\documentclass[11pt]{article}

\usepackage{empheq,fancybox}  % allows \begin{empheq}[box=\fbox]{align} XXXX \\ XXXX \end{empheq}
\usepackage{amsmath}	%Allows \begin{equation}, \b=egin{align}, \underbrace{2+2=4}_{\text{obvious}}
\usepackage{bm}		    %Allows {\bm \nabla \alpha} for bold symbol
\usepackage{bbm}
\usepackage{amssymb}	%Allows use math symbols
\usepackage{makeidx}	%Allows to use \makeindex and \index{key} and \printindex
\usepackage{color} 		%Allows {\color{red} Hello World} or \colorbox{blue}{Hello World}
\usepackage{graphicx} 	%Allows \includegraphics{figure.pdf}
\usepackage{enumitem}
\usepackage{multirow}	% Multirows in tables
\usepackage{times} 		%Use Times New Roman font
\usepackage{tensor} 	%Allows to use \tensor[^{\alpha}_{\beta}]{T}{^{\mu}_{\nu}}
\usepackage{hyperref} 	%Automatically links \label and \ref commands; Always load last
\usepackage{cite} 		%Allows \cite{ref}.  
\usepackage{slashed}
\usepackage[all]{hypcap}%Ensures figure hyperlinks are positioned to display entire figure
\usepackage[title]{appendix}
\usepackage{notoccite}
\usepackage[font=small]{caption}
\hypersetup{
    colorlinks=true,	%false: black links; true: colored links
    linkcolor=red,		%color of internal links
    citecolor=blue,		%color of links to bibliography
    filecolor=magenta,	%color of file links
    urlcolor=black	    %color of external links
}
\newcommand{\per}{\ . \ }
\newcommand{\com}{\ , \ }

\newcommand{\eg}{{\it e.g.}}
\newcommand{\xvec}{{\bm x}}

\newcommand{\kvec}{{\bm k}}

\newcommand{\khat}{\hat{\kvec}}
\newcommand{\kabs}{|\kvec|}

\newcommand{\HRH}{H_\mathrm{RH}}
\newcommand{\aRH}{a_\mathrm{RH}}

\newcommand{\TRH}{T_\mathrm{RH}}

\newcommand{\alphaRH}{\alpha_\mathrm{RH}}
\newcommand{\etaRH}{\eta_\mathrm{RH}}
\newcommand{\gRH}{g_{*\mathrm{RH}}}

\newcommand{\meff}{m_\mathrm{eff}}

\newcommand{\IdS}{\mathrm{I_{dS}}}
\newcommand{\IIdS}{\mathrm{II_{dS}}}
\newcommand{\IIIdS}{\mathrm{III_{dS}}}
\newcommand{\IVdS}{\mathrm{IV_{dS}}}
\newcommand{\IMD}{\mathrm{I_{MD}}}
\newcommand{\IIMD}{\mathrm{II_{MD}}}
\newcommand{\IIIMD}{\mathrm{III_{MD}}}
\newcommand{\IVMD}{\mathrm{IV_{MD}}}

\newcommand{\IRD}{\mathrm{I_{RD}}}

\newcommand{\IIIRD}{\mathrm{III_{RD}}}
\newcommand{\IVRD}{\mathrm{IV_{RD}}}

\newcommand{\eref}[1]{Eq.~(\ref{#1})}
\newcommand{\erefs}[2]{Eqs.~(\ref{#1})~and~(\ref{#2})}
\newcommand{\fref}[1]{Fig.~\ref{#1}}
\newcommand{\sref}[1]{Sec.~\ref{#1}}

\newcommand{\tref}[1]{Table~\ref{#1}}
\newcommand{\rref}[1]{Ref.~\cite{#1}}
\newcommand{\pref}[1]{(\ref{#1})}
\newcommand{\half}{{\textstyle\frac{1}{2}}}

\newcommand{\Lcal}{\mathcal{L}}
\newcommand{\GeV}{\ \mathrm{GeV}}
\newcommand{\MeV}{\ \mathrm{MeV}}
\newcommand{\eV}{\ \mathrm{eV}}

\newcommand{\cm}{\ \mathrm{cm}}

\newcommand{\Mpl}{M_{\rm Pl}}
\newcommand{\ket}[1]{\bigl|#1\bigr>}
\newcommand{\expval}[3]{\bigl< #1 \bigr| #2 \bigl| #3 \bigr>} %bra-ket

\newcommand{\nn}{\nonumber \\}
\numberwithin{equation}{section}
\begin{document}
%------------------------------ 
\title{Completely Dark Photons from \\ Gravitational Particle Production \\ During the Inflationary Era} 
\author{Edward W.\ Kolb$^1$ and Andrew J.\ Long$^2$}
\date{%
    $^1$Kavli Institute for Cosmological Physics and Enrico Fermi Institute\\  The University of Chicago, 5640 S.\ Ellis Ave., Chicago, IL 60637\\%
    $^2$Department of Physics and Astronomy, Rice University, Houston, Texas 77005\\[2ex]%
    \today
}
\maketitle
\begin{abstract} Starting with the de Broglie--Proca Lagrangian for a massive vector field, we calculate the number density of particles resulting from gravitational particle production (GPP) during inflation, with detailed consideration to the evolution of the number density through the reheating. We find plausible scenarios for the production of dark-photon dark matter of mass in a wide range, as low as a micro-electron volt to $10^{14}\GeV$.  Gravitational particle production does not depend on any coupling of the dark photon to standard-model particles. 

\end{abstract}
\begingroup
\setlength{\parskip}{-0.5ex}
\hypersetup{linkcolor=black}
\tableofcontents
\endgroup

%------------------------------ 
\section{Introduction}\label{sec:intro}
%------------------------------ 

An abundance of observational evidence indicates that our Universe is filled with a mysterious, invisible substance that we call dark matter \cite{Bertone:2016nfn}.    Assuming that the dark matter is a weakly-interacting collection of as-yet-unidentified elementary particles, the viable theory space is vast.  Apart from weak constraints on the dark matter particle's mass, $10^{-22}\eV \lesssim m \lesssim 10^{19}\GeV$ where the lower limit is from the requirement that the de Broglie wavelength of the particle is less than the size of dark-matter dominated objects and the upper limit is the requirement that the particle is not a black hole, the cosmological and astrophysical data provide no solid additional information about the dark matter's other properties (\textit{e.g.}, spin), and little information about the dark matter's interactions, apart from the fact that it must couple extremely weakly to visible matter.  In fact, it's useful to bear in mind that the data is consistent with a model of dark matter that only interacts gravitationally with visible-sector matter.  But if the dark-matter particle has only gravitational interactions with visible matter, the question arises: ``How was the dark matter produced in the early universe?'' A natural answer is that the origin of the dark matter must be through its gravitational interactions.  That is the explanation we pursue.   In this work we assume that the dark matter is a massive spin-1 particle, and we study the creation of dark matter during inflation and reheating through the phenomenon of gravitational particle production (GPP) in the inflationary era.\footnote{In this paper, by GPP we restrict ourselves to the phenomenon of gravitational production due to the nonadiabatic evolution of a field during inflation.  We do not consider other ``gravitational'' scenarios such as production from the standard-model plasma via graviton exchange \cite{Garny:2015sjg} or the misalignment mechanism~\cite{Nelson:2011sf,Arias:2012az,Nakayama:2019rhg}.  }

A massive and stable spin-1 particle, which is often called a dark photon, provides a viable candidate for the dark matter~\cite{Essig:2013lka, Fabbrichesi:2020wbt}.   If the dark photon couples non-gravitationally to visible matter, for instance through a gauge-kinetic mixing or because it is the force carrier for B$-$L, then there are possible mechanisms for early-Universe production.  However, if the dark photon is ultra-light, then interactions like kinetic mixing (alone) do not lead to dark-matter production in the early universe.  For instance, any dark photons produced from the plasma via thermal freeze-in or freeze-out would have energy $E \sim T_\mathrm{plasma}$ at their time of production and energy $E \sim T_\mathrm{cmb} \sim \mathrm{eV}$ at radiation-matter equality (assuming no entropy production that would lead to a higher plasma temperature).  For masses $m \lesssim \mathrm{eV}$ these particles would not be cold dark matter, but rather hot dark radiation~\cite{KolbTurner:1990,    Baumann:2016wac}.  This problem of ultra-light dark-photon production has attracted significant attention and model-building efforts lately~\cite{Co:2018lka,Agrawal:2018vin,Bastero-Gil:2018uel, Dror:2018pdh, Long:2019lwl, Nakai:2020cfw}, and it motivates us to consider dark-photon creation via GPP.  

The phenomenon of gravitational particle production \cite{Parker:1969au,Parker:1974qw,Fulling:1974zr,Ford:1986sy, Anderson:1987yt, Lyth:1996yj} results from the behavior of quantum fields in curved spacetime geometries \cite{DeWitt:1975ys, BirrellDavies:1982, Parker:2009uva}.  It has been studied in a variety of contexts, including most notably black holes (Hawking radiation) \cite{Hawking:1974rv} and cosmological inflation (inflationary quantum fluctuations) \cite{Mukhanov:1988jd, Sasaki:1983kd, Kodama:1985bj}.  In the context of dark matter, the gravitational production of spin-0 particles was studied by Refs.~\cite{Chung:1998zb,Kuzmin:1998uv,Giudice:1999yt,Ema:2018ucl, Markkanen:2018gcw, Hashiba:2018tbu, Fairbairn:2018bsw,Tenkanen:2019aij,Herring:2019hbe, Hashiba:2019mzm}, spin-1/2 particles by Refs.~\cite{Chung:2011ck,Ema:2019yrd, Herring:2020cah}, spin-1 particles by Refs.~\cite{Dimopoulos:2006ms,Graham:2015rva,Ema:2019yrd,Ahmed:2020fhc}, and spin-3/2 particles by Refs.~\cite{Kallosh:1999jj, Giudice:1999yt, Giudice:1999am}.  We discuss the physics of GPP in \sref{sec:GPP}.   At this point, it is worth remarking that GPP is a general consequence of quantum field theory and general relativity for any field (unless all operators involving the field are invariant under a Weyl conformal transformation).  In the case of dark matter, which must have a nonzero mass, the question is not whether gravitational production occurs, but rather how much dark matter is generated in this way.  

\begin{figure}[p]
\begin{center}
\includegraphics[width=0.95\textwidth]{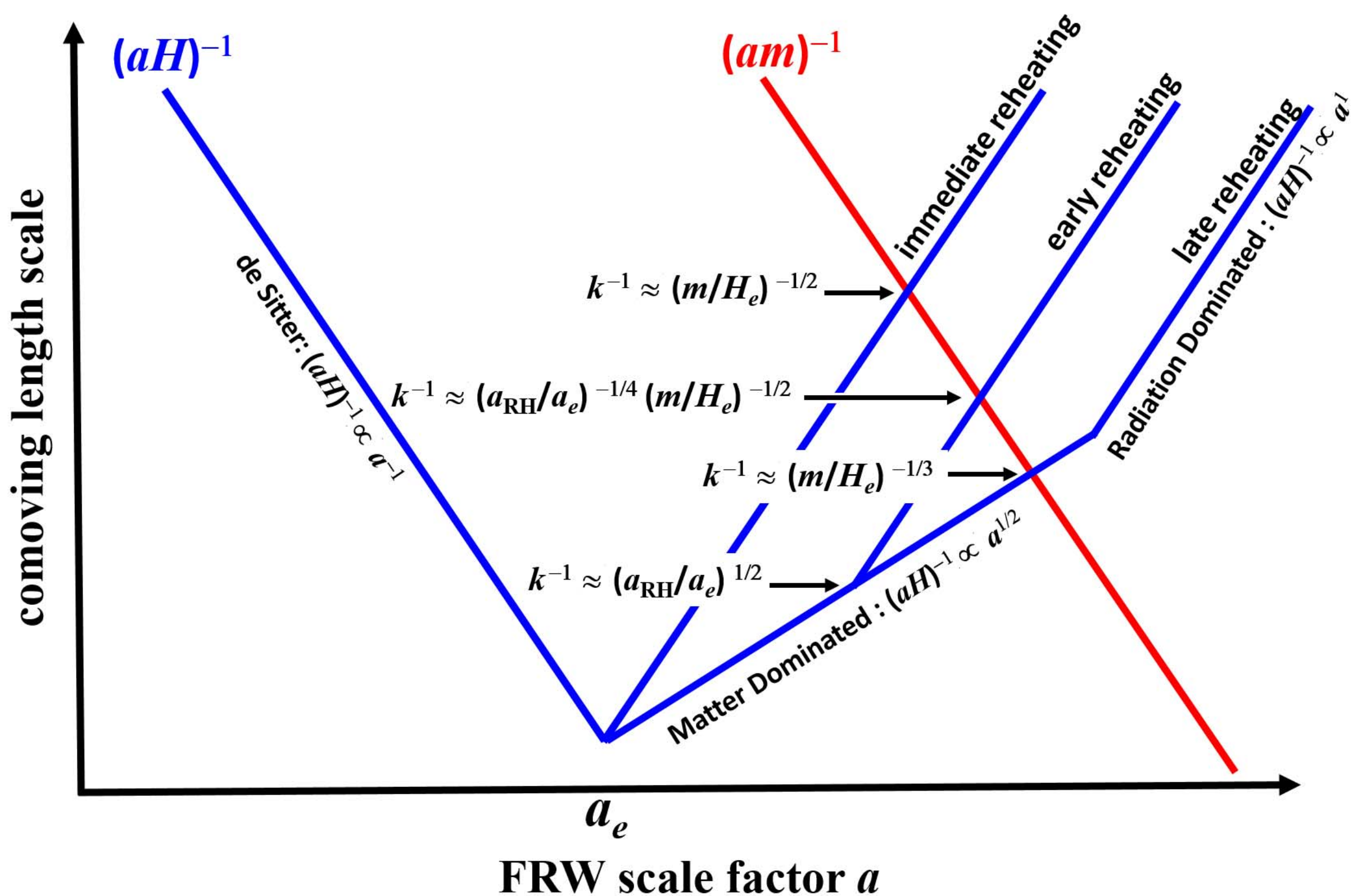}\\
\caption{This diagram illustrates the comoving Hubble scale $(aH)^{-1}$ and the comoving Compton wavelength $(am)^{-1}$ of the vector field for the three reheating scenarios discussed in the text, and it highlights the comoving wavenumbers, $k = 2\pi / \lambda$, that are important for understanding the spectrum of gravitationally-produced spin-1 dark matter.  Inflation ends at $a = a_e$ when $H(a_e) = H_e$, and we assume $m \ll H_e$.  In the de Sitter phase $H\approx\mathrm{const.}$, before reheating $H\propto a^{-3/2}$, and after reheating $H\propto a^{-2}$.  We illustrate three possibilities: \textit{Late Reheating:}  If reheating has not completed by the time when $m \approx H(a)$, then the spectrum is well-approximated by a power law with a single break at $k^{-1} \approx (m / H_e)^{-1/3}$, which corresponds to the special mode that reenters the Hubble radius at the same time when $m = H$.   \textit{Early Reheating:}  If reheating completes before $m \approx H(a)$ then the spectrum is a power law with two breaks.  \textit{Immediate Reheating:}  In the limit where the duration of reheating goes to zero, the early reheating scenario is approximated by the immediate reheating scenario in which the spectrum is a power law with its break at $k^{-1} \approx (m/H_e)^{-1/2}$.  \label{fig:inverse}}
\end{center}
\end{figure}

It was realized by Graham, Mardon, and Rajendran \cite{Graham:2015rva} (hereafter GMR) that dark-photon dark matter could be produced gravitationally and that the correct relic abundance could be obtained for masses as low as $m \gtrsim 10^{-6} \eV$.  The analysis in GMR assumed that reheating occurred immediately after the end of inflation, so that the universe immediately transitioned from a quasi-de Sitter phase of inflation into a radiation-dominated era.  Even though reheating is never truly instantaneous, for ultra-light dark-photon dark matter this is a reasonable assumption, since the spectrum and relic abundance are insensitive to the reheating history as long as reheating completes sufficiently early (before $H(t) = m$).  In this work, we extend the original analysis of GMR to account for the finite duration of reheating, which is assumed to be a phase of matter domination.  The diagram in \fref{fig:inverse} summarizes our model for the spacetime geometry during reheating, and anticipates how the spectrum of dark matter depends on the reheating history.  We find that the spectrum of gravitationally-produced spin-1 particles is modified for masses $m \gtrsim (1 \GeV) (T_\mathrm{RH} / 10^9 \GeV)^2$, and it takes the form of a broken power law with two breaks.  The total relic abundance (integral of the spectrum) is shown to be relatively insensitive to the reheating history for ultralight dark photons.  

Our study is closely related to the work that's presented in \rref{Ahmed:2020fhc}.  The authors of that article have also studied the gravitational production of spin-1 dark matter while accounting for the finite duration of reheating.  We follow a similar analysis here, but with two notable differences in our assumptions.  First, for our analytical calculations, we restrict our attention to models of reheating with equation of state $w = 0$, whereas the work in \rref{Ahmed:2020fhc} allows for a more general range $-1/3<w<1$.  Our assumption is motivated by models of inflation with a quadratic inflaton potential near the minimum, which predict $w = 0$.  Our results generally agree with the $w=0$ case in \rref{Ahmed:2020fhc}; \textit{e.g.}, compare our \tref{table:finalsummary} with their Eq.~(3.31).  Second, for our numerical calculations, we study a quadratic inflaton potential, $V(\phi) = m_\phi^2 \phi^2 / 2$, and we solve the inflaton's equation of motion to determine the background spacetime, \textit{i.e.} $a(t)$, $H(t)$, and $R(t)$.  By contrast, \rref{Ahmed:2020fhc} assumes an exact de Sitter phase of inflation followed by an immediate transition into reheating with equation of state $-1/3<w<1$.  As we show in \fref{fig:totalspectrum}, accounting for the evolution of $H(t)$ during inflation, as we have done here, can lead to an  $O(10)$ change in the predicted dark matter relic abundance.   

The reader should also compare our work with \rref{Ema:2019yrd}, in which the authors present a systematic study of gravitational particle production for vector dark matter (and also spin-1/2 fermions).  The authors of \rref{Ema:2019yrd} recognize that gravitational particle production can be efficient even for particles with mass above the inflationary Hubble scale but below the inflaton mass scale, $H_\mathrm{inf} < m < m_\phi$.  Our work focuses instead on light dark matter with $m \ll H_\mathrm{inf} \sim m_\phi$.  Our analytic results generally agree with the light vector boson case of \rref{Ema:2019yrd}, which is also in agreement with the earlier \rref{Graham:2015rva}.  

%============
%============

For those interested in the final answer, the result for the contribution to the present mass density of dark matter, parameterized by $\Omega h^2/0.12$, is shown in \tref{table:finalsummary}.  In the table $\TRH$ is the reheat temperature (and $\TRH^\mathrm{MAX}$ is its maximum possible value) discussed in \sref{sec:FRW}, $\aRH/a_e$ is the ratio of the scale factor at reheating to the scale factor at the end of inflation, $m$ is the mass of the dark photon, and $H_e$ is the expansion rate of the Universe at the end of inflation.

%===================
\begin{table}[t]
\begin{center}
\caption{\label{table:finalsummary} Results for $\Omega h^2/0.12$ for immediate, early, and late reheating. The expression for $\TRH^\mathrm{MAX}$ is derived in the discussion before \eref{eq:TRHETC}. The lower limit to $\TRH$ is due to the requirement that the RD universe is able to produce the neutrino background \protect\cite{PhysRevD.92.123534}.}
\begin{tabular}{|c|c|}\hline
Early \& Immediate Reheating & Late Reheating\\ 
$(m/H_e)^{-2/3}>\aRH/a_e\geq1$ & $\aRH/a_e >(m/H_e)^{-2/3}$ \\ [1ex]
$\TRH^\mathrm{MAX}> \TRH > 8.4\times10^8\left(\dfrac{m}{\GeV}\right)^{1/2} \GeV$ & $8.4\times10^8\left(\dfrac{m}{\GeV}\right)^{1/2} \GeV > \TRH > 4.7\MeV $ \\[2ex]\hline 
& \\
$\dfrac{\Omega h^2}{0.12} = \left(\dfrac{m}{10^{-6}\eV}\right)^{1/2}\ \left(\dfrac{H_e}{10^{14}\GeV}\right)^2$ 
& $\dfrac{\Omega h^2}{0.12} = \left(\dfrac{\TRH}{5\times10^7\GeV}\right)\ \left(\dfrac{H_e}{10^{11}\GeV}\right)^2$ \\[3ex] \hline
\end{tabular}
\end{center}
\end{table}
%===================

The remainder of this article is organized as follows.  We present the massive vector model in \sref{sec:model} for a Minkowski spacetime background, and we extend it to an inflationary background in \sref{sec:FRW} before reviewing the phenomenon of gravitational particle production in \sref{sec:GPP}.  Our main results appear in Sec.~\ref{sec:density}~and~\ref{sec:relic}, where we solve the vector field's mode equations -- both analytically and numerically -- to calculate the spectrum and relic abundance of gravitationally-produced spin-1 dark-matter particles.  We summarize and conclude in \sref{sec:conc}.

%------------------------------ 
\section{The massive vector model \label{sec:model}}
%------------------------------ 

We will only consider spin-1 fields with non-zero mass because, as we shall see, massless spin-1 fields (e.g., electrodynamics) are conformally coupled to gravity and will not be produced by the expansion of the universe.  Since we are interested in GPP of massive vectors as a source of dark matter, our considerations will not apply to the massive spin-1 particles of the standard model ($W^\pm$ and $Z$).  The vector field must transform as the $(\half,\half)$ representation of the Lorentz group.  It contains components with helicity $1$ and $0$. 

For the analysis of massive spin-1 bosons, we start with the de Broglie-Proca action in Minkowski space \cite{deBroglie:1922, deBroglie:1934, Proca:1900nv}:
\begin{align}\label{eq:deBroglieProca}
S=\int d^4x \left(-\tfrac{1}{4}\eta^{\mu\alpha}\eta^{\nu\beta}F_{\alpha\beta}F_{\mu\nu} + \half m^2\eta^{\mu\nu}A_\mu A_\nu \right) \per
\end{align}
Here $\eta^{\mu\nu} = \mathrm{diag}(1,-1,-1,-1)$ is the Minkowski metric and $F_{\mu\nu}$ is the field strength tensor.  We will see that in the massive (as in the massless) theory, $A^0$ is not dynamical.  This Lagrangian is the unique renormalizable Lorentz-invariant Lagrangian for a massive spin-1 field.\footnote{In this section we follow Weinberg \cite{Weinberg:1995mt}.} 
 
Unlike the familiar electroweak theory, the de Broglie-Proca Lagrangian does not describe a gauge theory because the mass term explicitly breaks gauge invariance, i.e., invariance under the local transformation $A_\mu(x) \rightarrow A_\mu(x) + \partial_\mu\alpha(x)$.  However, we can view the action of \eref{eq:deBroglieProca} as the effective low-energy theory of a gauge theory, namely the Abelian-Higgs model with a complex scalar field $\Phi$ which obtains a vacuum expectation value $v$.  Assuming $D_\mu\Phi = \partial_\mu\Phi - i g A_\mu\Phi$, where $A_\mu$ is a \textit{massless} gauge field, after symmetry breaking and integrating out the massive scalar, the effective theory is equivalent to the de Broglie-Proca theory with the mass of the vector field  $m=g v$.  In this approach the de Broglie-Proca Lagrangian is the effective low-energy theory of an Abelian-Higgs model in the limit $v \rightarrow \infty$, $g \rightarrow 0$, and $gv \rightarrow \mathrm{const}$.   We could relax the $v\to\infty$, $g\to0$ limit and just assume an Abelian Higgs model where the mass of the Higgs (of course this is not the electroweak Higgs) is larger than $H$ during inflation while the mass of the vector is of order or smaller than the expansion rate during inflation.  Or perhaps the Higgs is produced during inflation and then decays.  It would presumably decay to the massive vector, so there would be two sources of remnant vectors: GPP of the massive field during inflation, and production of the massive vector through Higgs decay.\footnote{If the massive spin-1 dark photon arises from an Abelian Higgs model in the UV, then the theory predicts an additional spin-0 Higgs boson.  We would have to assume that its mass is larger than $2m$ so that it is unstable and decays pairwise into dark photons.  Additionally, we would have to assume that its mass is larger than $\mathcal{O}(\mathrm{few} \times m_\mathrm{inflaton})$ so that its gravitational production is suppressed.} 

The antisymmetric field-strength tensor in terms of the vector field $A_\mu$ is given by
\begin{align}
F_{\mu\nu}= \partial_\mu A_\nu - \partial_\nu A_\mu \per
\end{align} 
The classical equation of motion is the so-called Proca equation
\begin{equation}
\partial_\mu F^{\mu\nu} + m^2A^\nu = 0 \per
\end{equation}
Note that since $\partial_\nu\partial_\mu F^{\mu\nu}= 0$, from the Proca equation we find the Lorenz gauge condition $\partial_\nu A^\nu=0$.  This condition, usually set by gauge fixing in the massless theory, is a consequence of the equation of motion of the massive theory. From the Proca equation, the gauge field satisfies four copies of the Klein-Gordon equation for the four components of $A_\mu$:
\begin{equation}
\eta^{\alpha\beta}\partial_\alpha\partial_\beta A^\mu +m^2 A^\mu = 0 \per 
\end{equation}
The conjugate momenta to $A^\mu$ are $\pi_\mu=\partial\Lcal/\partial \dot{A}^\mu = F_{0\mu}$.  Unlike the massless vector case, the fact that $\pi_0=0$ will not be a problem because $A^0$ will be an auxiliary field.

Unlike electrodynamics, which has two physical (transverse) degrees of freedom, for the massive theory there are three degrees of freedom, namely two transverse degrees of freedom, which will be denoted by $A^T$, and one longitudinal degree of freedom, which will be denoted by $A^L$.  The $m\rightarrow 0$ limit is tricky.  The longitudinal mode survives in the $m\rightarrow 0$ limit, but it is decoupled from the other degrees of freedom and behaves like a scalar degree of freedom (the Goldstone boson equivalence theorem).   

In component form the action is\footnote{In this section we follow the analysis of Graham, et al., \cite{Graham:2015rva}.}
\begin{align}\label{eq:actionconponents}
S\left[A_\mu(t,\xvec)\right] = \int d^4x\sqrt{-\eta}\left[ \half \left(\partial_tA_i-\partial_iA_t\right)^2 - \tfrac{1}{4} \left(\partial_iA_j-\partial_jA_i\right)^2 +\half m^2A_t^2 -\half m^2 A_i^2\right] 
\end{align}
where $\sqrt{-\eta}=1$ is the determinant of the Minkowski metric.  
Note that $A_t$ does not have a kinetic term; it is an auxiliary field.  The field equations in component form are
\begin{align}\label{eq:eomcomponents}
\left[ \delta_{ij}\partial_t^2 - \left(\delta_{ij}\partial_k^2-\partial_i\partial_j\right) + \delta_{ij}m^2\right]A_j - \partial_i\partial_tA_t & = 0 \nn
\left(\partial_j^2 - m^2\right)A_t - \partial_t\partial_jA_j & = 0 \per
\end{align} 

Since $A^\mu$ satisfies the Klein-Gordon equation we can again expand it as
\begin{align}\label{eq:vecfour}
A^\mu(t,\xvec) = \int \frac{d^3\kvec}{(2\pi^3)}\ A_\kvec^\mu(t)\ e^{i\kvec\cdot\xvec} \per
\end{align}
In terms of the normal modes the action \pref{eq:actionconponents} becomes
\begin{align}\label{eq:actionfourier}
S\left[A_\mu(t,\xvec)\right] & = \int dt\ \int\frac{d^3\kvec}{(2\pi^3)}\left[ \tfrac{i}{2}\kvec_iA^*_t(\partial_tA_i) - \tfrac{i}{2}\kvec_i(\partial_tA^*_i)A_t  + \half \left(\kabs^2+m^2\right)\left|A_t\right|^2 \right. \nn
& \left. - \tfrac{1}{4} \left|\kvec_iA_j-\kvec_j A_i\right|^2  + \half\left|\partial_tA_i\right|^2 - \half m^2\left|A_i\right|^2  \right] \com
\end{align}
where in the interest of notational simplicity we have suppressed the $\kvec$ label on $A_i$ and $A_t$ inside the integral.  In order to solve for the temporal component of the field we rewrite \eref{eq:actionfourier} as 
\begin{align}\label{eq:actionfourier2}
S\left[A_\mu(t,\xvec)\right] & =  \int dt\ \int\frac{d^3\kvec}{(2\pi^3)}\Biggl[ \half (\kabs^2+m^2) \left|A_t+i\dfrac{\kvec_i(\partial_tA_i)}{\kabs^2+m^2}\right|^2 - \half \frac{\left|\kvec_i(\partial_tA_i)\right|^2}{\kabs^2+m^2} \nn
& - \tfrac{1}{4} \left|\kvec_iA_j-\kvec_j A_i\right|^2 + \half\left|\partial_tA_i\right|^2 -\half m^2\left|A_i\right|^2
\Biggr] \per
\end{align}
Now that $A_t$ is isolated it is clear it is nondynamical and we can solve for it:
\begin{align}\label{eq:integrateout}
A_t = - i\dfrac{\kvec_i(\partial_tA_i)}{\kabs^2+m^2}.
\end{align}
After integrating out $A_t$ the action becomes
\begin{align}\label{eq:abecomes}
S\left[A_\mu(t,\xvec)\right] & =  \int dt\ \int\frac{d^3\kvec}{(2\pi^3)}\biggl\{ \half (\partial_tA_i^*)\left(\delta_{ij}-\dfrac{\kvec_i\kvec_j}{\kabs^2+m^2}\right)(\partial_tA_j) \nn
& - \half A_i^*\left[(\kabs^2+m^2)\delta_{ij}-\kvec_i\kvec_j\right]A_j
\biggr\} \per
\end{align}

Now it is useful to further decompose the spatial components of the vector field into transverse and longitudinal polarization modes.  This is accomplished by first writing $ A_i = {\bm A}_{\kvec i}$ where ${\bm A}_{\kvec}(x^0,\kvec)$ is a complex 3-vector.  
Note that the mapping from 4-vector to 3-vector is performed using the \textit{covariant 4-vector} with a lowered index.  We then decompose the 3-vector as 
\begin{align}
{\bm A}_\kvec = A^{T_1}_\kvec \, {\bm \varepsilon}^{T1} + A^{T_2}_\kvec \, {\bm \varepsilon}^{T_2} + A^{L}_\kvec \, {\bm \varepsilon}^{L}
\end{align}
where $A^{T_1}_\kvec$, $A^{T_2}_\kvec$, and $A^{L}_\kvec$ are complex mode functions for the two transverse and the single longitudinal polarization mode, and ${\bm \varepsilon}^{T_1}(\khat)$, ${\bm \varepsilon}^{T_2}(\khat)$, and ${\bm \varepsilon}^{L}(\khat)$ are the polarization vectors, which satisfy 
\begin{align}\label{eq:varepsilon}
	& {\bm \varepsilon}^{T_1} \cdot {\bm \varepsilon}^{T_1} = {\bm \varepsilon}^{T_2} \cdot {\bm \varepsilon}^{T_2} = {\bm \varepsilon}^{L} \cdot {\bm \varepsilon}^{L} = 1 \nn
	& {\bm \varepsilon}^{T_1} \cdot {\bm \varepsilon}^{T_2} = {\bm \varepsilon}^{T_1} \cdot {\bm \varepsilon}^{L} = {\bm \varepsilon}^{T_2} \cdot {\bm \varepsilon}^{L} = 0 \nn
	& {\bm \varepsilon}^{L} = \khat 
	\per
\end{align}
Then the action can be broken into two terms,
\begin{align}
S[A_\kvec^{T_1}, A_\kvec^{T_2}, A_\kvec^L] = S^T[A_\kvec^{T_1}, A_\kvec^{T_2}] + S^L[A_\kvec^L] 	\com
\end{align}  
where
\begin{subequations} \label{eq:RV_action_6} \begin{align}
S^T & = \sum_{b=1,2} \int dt \, \int \! \! \frac{d^3\kvec}{(2\pi)^3} \, \left[ \half  |\partial_t A_\kvec^{T_b}|^2 - \half \left(\kabs^2 + m^2 \right) \ |A_\kvec^{T_b}|^2\right] \\ 
S^L  & = \int dt \, \int \! \! \frac{d^3\kvec}{(2\pi)^3} \, \left[ \half \frac{m^2}{\kabs^2 + m^2} \, |\partial_t A_\kvec^L|^2 - \half m^2 |A_\kvec^L|^2 \right] 	\per
\end{align} \end{subequations}
Note also that the long-wavelength modes for which $\kabs \to 0$ behave identically for the two transverse polarizations and the longitudinal polarization.  

Although the kinetic term for $A_\kvec^{T_b}$ is canonically normalized, the kinetic term for $A_\kvec^L$ is not.  Therefore we define the field $\phi^L$ via 
\begin{align}\label{eq:alphil}
A^L_\kvec = \sqrt{\frac{k^2+m^2}{m^2}}\ \phi^L_\kvec \per
\end{align}
In terms of $\phi^L_\kvec$, the action for the longitudinal mode is
\begin{align}
S^L & =  \int dt \, \int \! \! \frac{d^3\kvec}{(2\pi)^3} \, \Bigl[ \half    |\partial_t \phi_\kvec^L|^2 - \half \left(\kabs^2 + m^2 \right) \ |\phi^L_\kvec|\Bigr] \per
\end{align}

After much manipulation we ended up with the action for two scalars, $A^T$ and $\phi^L$ ($A^T$ has two degrees of freedom).  Although in Minkowski space the action ends up being just the action for scalars, in a curved spacetime the result won't be quite so simple.

Using the Belifante--Rosenfeld stress-energy tensor and \eref{eq:deBroglieProca} for $\Lcal$, we find
\begin{align}
T^{\mu\nu} & = \frac{1}{4} \left( \eta^{\mu\nu} \eta^{\alpha\gamma} \eta^{\beta\delta} - 4 \eta^{\mu\alpha} \eta^{\nu\gamma} \eta^{\beta\delta} \right) F_{\alpha\beta} F_{\gamma\delta} + m^2  \left( \eta^{\mu\alpha} \eta^{\nu\beta} - \half  \eta^{\mu\nu} \eta^{\alpha\beta} \right) A_\alpha A_\beta \per
\end{align}
This yields $\rho=T_{00}$ as
\begin{align}\label{eq:rho1mink}
\rho & = \frac{1}{4} \eta^{\alpha\gamma} \eta^{\beta\delta} F_{\alpha\beta} F_{\gamma\delta} - \eta^{\beta\delta}F_{0\beta}F_{0\delta} 	+ m^2A_t^2 - \half m^2 \eta^{\alpha\beta}A_\alpha A_\beta 	\nn 
& = \half(\partial_tA_i-\partial_iA_t)^2 + \tfrac{1}{4}(\partial_iA_j-\partial_jA_i)^2 + m^2A_t^2 + m^2A_i^2 \per
\end{align}

%------------------------------ 
\section{de Broglie--Proca in a Friedmann--Robertson--Walker background \label{sec:FRW}}
%------------------------------ 

Before proceeding we have to specify a background geometry.  We will consider the action of \eref{eq:deBroglieProca} in a particular curved space, namely the Friedmann--Robertson--Walker (FRW) spacetime.  Since we are concerned with the early-universe evolution we are justified in taking the spatially-flat FRW metric $ds^2=dt^2-a^2(t)d\xvec^2$.\footnote{We adopt the Landau-Lifshitz timelike conventions \cite{LL} for the signature of the metric ($\mathrm{sign}[\eta_{00}]=+1$ where $\eta_{\mu\nu}$ is the Minkowski metric), the Riemann curvature tensor  ($\tensor{R}{^\rho_{\sigma\mu\nu}} =+ \partial_\mu \Gamma^\rho_{\nu\sigma} \cdots$), and the sign of the Einstein tensor $G_{\mu\nu}=+8\pi G_N T_{\mu\nu}$. To translate these conventions to other conventions, see the introductory material in Misner, Thorne, and Wheeler \cite{MTW}. Our sign conventions correspond to $(-,+,+)$ in their table.}  In conformal time $\eta$ the metric is simply $ds^2=a^2(\eta)\left(d\eta^2-d\xvec^2\right)$. We will assume an initial inflationary epoch terminating at $a=a_e$, followed by a matter-dominated (MD) era that ends with reheating at $a=\aRH$.\footnote{When we refer to the values of quantities at reheating, we mean the values when the universe becomes radiation dominated after inflation.}

We choose $a$ to have dimension of length (hence, coordinates $\eta$ and $\xvec$ are dimensionless).  In the spatially-flat case we are free to scale $a$.  We define $a_e$ to be the scale factor at the end of inflation.  Since we can set the scale, a convenient choice is $a_e=H_e^{-1}$ where $H_e$ is the expansion rate at the end of inflation.  Thus, 
\begin{align}\label{eq:convention}
a_eH_e=1 \per 
\end{align}

Since only $d\eta$ is significant, we are free to add or subtract anything to $\eta$.  A convenient choice is $\eta=0$ at the end of inflation.  Thus, $-\infty < \eta < +\infty$, with $\eta=0$ at the end of inflation.

We define the wavenumber of a Fourier mode, $k$, to be dimensionless.  The physical wavenumber with units of length$^{-1}$ is $k/a$. The physical wavenumber at the end of inflation is $k/a_e$.  Equating $k/a_e$ and $H_e$: $k/a_e=H_e$ gives $k=1$ for the wavenumber crossing the Hubble radius at the end of inflation (since $a_eH_e=1$).  

Finally, it is useful to define dimensionless parameters 
\begin{align}\label{eq:note}
\alpha \equiv \frac{a}{a_e}\  ; \qquad \mu \equiv \frac{m}{H_e}\ ; \qquad h \equiv \frac{H}{H_e}\ ; \qquad \alphaRH \equiv \frac{\aRH}{a_e} \per 
\end{align}
At the end of inflation and the beginning of the matter-dominated era, $\alpha=1$ and $h=1$.  At the end of the MD era and beginning of the RD era, $\alpha=\alphaRH$ and $h=h_\mathrm{RH}$.

For analytic work we will assume an initial exact de Sitter (dS) phase, followed by an immediate transition to a Matter-Dominated (MD) phase at $a=a_e$, followed by another immediate transition to a Radiation-Dominated (RD) phase at $a=\aRH$.  It will prove useful to collect the dependence of $\alpha$, $h$, and $R/6H_e^2$ on $\eta$ for the dS, MD, and RD eras together in a single place: \tref{table:aetaR}.

%===================
\begin{table}[ht]
\begin{center}
\caption{\label{table:aetaR} The dependence of the scale factor, the expansion rate, and the scalar curvature on conformal time $\eta$.  We assume that the de Sitter era is followed by a matter-dominated era until reheating, which initiates a radiation-dominated era.  Dimensionless variables are defined by Eqs.~(\ref{eq:convention})~and~(\ref{eq:note}).}
\begin{tabular}{|c|c|c|c|}\hline
& de Sitter (dS) & Matter-Dominated (MD) & Radiation-Dominated (RD) \\ 
& $0 < \alpha < 1$ & $1 < \alpha < \alphaRH$ & $\alphaRH<\alpha<\infty$ \\
& $-\infty<\eta<0$ & $0<\eta<\eta_\mathrm{RH}$ & $\etaRH<\eta<+\infty$ \\ \hline	
& & & \\
$\alpha$ & $\dfrac{1}{1-\eta}$ & $(1+\half\eta)^2$ & $\alphaRH\left[1+\alphaRH^{-1/2}(\eta-\eta_\mathrm{RH})\right]$ \\
&  & & \\
$\alpha$ & $\to-\dfrac{1}{\eta}\ \ (\eta\ll0)$ & $\to\dfrac{1}{4}\eta^2\ \ (0\ll\eta<\etaRH)$ & $\to \alphaRH^{1/2}\eta\ \ (\etaRH\ll\eta<+\infty)$ 
\\[2ex] \hline & & & \\ 
$h$ &  & $\dfrac{1}{(1+\half\eta)^3}$ & $\dfrac{\alphaRH^{-3/2}}{\left[1+\alphaRH^{-1/2}(\eta-\eta_{RH})\right]^2}$ \\
& $1$ &&\\
$h$ & & $\to\dfrac{8}{\eta^3} \ \ (0\ll\eta<\eta_\mathrm{RH})$ & $\to\dfrac{\alphaRH^{-1/2}}{\eta^2}\ \ (\etaRH\ll\eta<+\infty)$ \\[2ex]
\hline & & & \\
$\dfrac{R}{6H_e^2}$ &  & $-\dfrac{1}{2}\dfrac{1}{(1+\half\eta)^6}$ & 
\\ 
& $-2$ & & $0$ \\
& & & \\
$\dfrac{R}{6H_e^2}$ & & $\to\dfrac{32}{\eta^6} \ \ (0\ll\eta<\eta_\mathrm{RH})$ & \\[2ex]
\hline
\end{tabular}
\end{center}
\end{table}
%===================
 
For numerical results we will assume a chaotic inflation model.  In chaotic inflation the dynamics of inflation is determined by the dynamics of a scalar field known as the \textit{inflaton}. The inflaton potential is taken to be $V=\half m_\phi^2\phi^2$, where $m_\phi$ is the inflaton mass.  To be sure, this model is observationally challenged by precision CMB observations (see, e.g., \rref{Aghanim:2018eyx}), but it should serve our purposes and represent a large (but not exhaustive) class of slow-roll inflation models.  The end of inflation for this model occurred when $\phi \simeq 2.5\times10^{18}\GeV$, or roughly the reduced Planck mass.  The expansion rate at the end of inflation is $H_e\simeq m_\phi/2$.\footnote{We note that in the  chaotic model of inflation the inflaton mass and $H_e$ are approximately the same, but in general they can differ.  For example, hybrid or hilltop models allow $H_\mathrm{inf} \ll m_\phi$.  If they are very different, then the exponential suppression in GPP for $m>H_e$ can be avoided for $H_\mathrm{inf} \ll m_\mathrm{spectator} \ll m_\mathrm{inflaton}$ \cite{Ema:2016hlw,Chung:2018ayg}.}  In the simple single-field model of inflation the expansion rate during inflation is related to the amplitude of gravitational waves produced during inflation.  The present limit on the gravitational wave contribution to the CMB limits $H$ to be $H\lesssim 7.5\times10^{13}\GeV$.  This is the limit on $H$ approximately 30-60 e-folds in $a$ before the end of inflation.  Thirty e-folds in scale factor before the end of inflation in this model corresponds to about 4 times $H_e$.  Therefore, the limit on $H_e$ is approximately $H_e \lesssim 3\times10^{14}\GeV$. We will display the dependence on $H_e$. 

After the end of inflation in the chaotic model the field reaches the minimum of the potential and commences oscillations about the minimum of the potential.  During this period of oscillation about the minimum of the potential the amplitude of oscillations decreases due to the $-3H\dot{\phi}$ term in the equation of motion.  In the oscillatory phase $\dot{\rho}_\phi+3H\dot{\phi}^2=0$. Since $\phi$ rapidly (compared to $H$) oscillates about the minimum of the potential, $\dot{\phi}$ can be replaced by its average over an oscillation cycle, $\langle\dot{\phi}^2 \rangle_\mathrm{cycle} = \rho_\phi$, and  $\dot{\rho}_\phi+3H\rho_\phi=0$, exactly the behavior of a matter-dominated universe.  Of course the oscillatory phase cannot continue indefinitely.  The $\phi$ field must eventually decay into radiation.  This can be modeled by including in the equation of motion a decay term $\Gamma_\phi\dot{\phi}$.  If $\Gamma_\phi \ll H_e$, the additional term will only be important during the oscillatory phase.  Because of the $\Gamma_\phi$ term the coherent energy in the $\phi$ oscillations are converted to light degrees of freedom (radiation) and the universe ``reheats.''\footnote{``Reheat'' is somewhat of a misnomer since $\TRH$ is not the maximum temperature reached after inflation: see \eg, \rref{Giudice:2000ex}.}  The temperature of the universe when it becomes radiation dominated is known as the reheat temperature, $\TRH$.  We will display the dependence on $\TRH$.
 
Not much is known about the reheat temperature.  Clearly the universe was radiation dominated during big-bang nucleosynthesis, so a reasonable lower bound on $\TRH$ might be a few MeV.  In order to thermalize the neutrino background (as detected in the CMB) the reheat temperature must be greater than $\TRH>4.7\MeV$ \cite{PhysRevD.92.123534}.  If all of the inflaton energy density is immediately converted to radiation at reheating, then $(\pi^2/30)\gRH \TRH^4=3\HRH^2\Mpl^2$.  Here $\gRH$ counts the effective number of degrees of freedom in the radiation at a temperature of $\TRH$.  We will set $\gRH=106.75$, the value counting the number of effective degrees of freedom in the standard model.  Since there are orders on magnitude uncertainty in $H_e$ and $\TRH$ we will not bother carrying the dependence on $\gRH$. For immediate reheating, $\HRH=H_e$, and $\TRH^\mathrm{MAX}/10^9\GeV=8.4\times10^5(H_e/10^{12}\GeV)^{1/2}$.  This is the upper bound on $\TRH$. Using the fact that during the matter-dominated phase $H\propto a^{-3/2}$, then $\HRH/H_e = (a_e/\aRH)^{3/2}$, and we can relate $\TRH$, $H_e$, and $\aRH/a_e$:
\begin{align}\label{eq:TRHETC}
\alphaRH & = \left(\frac{90}{\pi^2g_*}\right)^{1/3} \frac{H_e^{2/3}\Mpl^{2/3}}{\TRH^{4/3}} =  8.0\times10^7\left(\frac{H_e}{10^{12}\GeV}\right)^{2/3}   \left(\frac{10^9\GeV}{\TRH}\right)^{4/3}  \nn
\frac{\TRH}{10^9\GeV} & = 8.4\times10^5\left(\frac{H_e}{10^{12}\GeV}\right)^{1/2}\alphaRH^{-3/4} \per 
\end{align} 
We will show below that the requirement for reheating to affect the final number density is that $\alphaRH<\mu^{-2/3}$.  So we will have to take evolution through the radiation-dominated era into account in calculating the final value of the number density if
\begin{align}\label{eq:MDRD}
\TRH < 8.4\times10^8\left(\frac{m}{\GeV}\right)^{1/2} \GeV \per
\end{align}

Promoting the action of \eref{eq:deBroglieProca} to a general spacetime with metric $g_{\mu\nu}(x)$ yields
\begin{align}\label{eq:RV_action_1}
S[A_\mu(x),g_{\mu\nu}(x)] = & \int \! d^4x \, \sqrt{-g} \left[ - \tfrac{1}{4} g^{\mu \alpha} g^{\nu \beta} F_{\mu \nu} F_{\alpha \beta} + \half m^2 g^{\mu \nu} A_{\mu} A_{\nu} \right. \nn
 & \left. - \half \xi_1 R g^{\mu\nu} A_\mu A_\nu - \half \xi_2 R^{\mu\nu}A_\mu A_\nu  \right] \per
\end{align}
The tensor structure of the vector field admits two different forms of dimension-4 operators describing non-minimal interactions of the vector field with the gravitational field, here proportional to the two constants $\xi_1$ and $\xi_2$. In our analysis eventually we will only consider minimal coupling $(\xi_1=\xi_2=0$), but we will carry the nonminimal terms to serve as a reference for possible future investigations.  The field strength tensor is $F_{\mu\nu} = \nabla_\mu A_\nu - \nabla_\nu A_\mu = \partial_\mu A_\nu - \partial_\nu A_\mu$ since the connection terms cancel.  Since we do not have to calculate loops, here we have neglected the gauge-fixing and ghost terms; see Eqs.\ (3.182) and (3.183) of Birrell \& Davies \cite{BirrellDavies:1982}. 

The equation of motion yields
\begin{align}\label{eq:RV_field_eqn_1a}
& \frac{1}{\sqrt{-g}} \partial_\mu \left[ \sqrt{-g} g^{\mu\alpha}g^{\nu\beta} F_{\alpha\beta} \right] + \left(m^2g^{\nu\beta} - \xi_1 Rg^{\nu\beta} - \xi_2 R^{\nu\beta} \right)A_\beta = 0 \per
\end{align}
The stress-energy tensor is 
\begin{align}\label{eq:RV_Tmn_1}
T^{\mu\nu} = & \frac{1}{4} \left( g^{\mu\nu} g^{\alpha\gamma} g^{\beta\delta} - 4 g^{\mu\alpha} g^{\nu\gamma} g^{\beta \delta} \right) F_{\alpha\beta} F_{\gamma\delta} + \left[ m^2 \left( g^{\mu\alpha} g^{\nu\beta} - \half g^{\mu\nu} g^{\alpha\beta} \right) - \xi_1 \left( R g^{\mu\alpha} g^{\nu\beta} + G^{\mu\nu} g^{\alpha\beta} \right) \right. \nn
& \left. - \xi_2 \left( g^{\mu\alpha} R^{\nu\beta} + g^{\mu\beta} R^{\nu\alpha} - \half g^{\mu\nu} R^{\alpha\beta} \right) \right] A_\alpha A_\beta + \left[ \xi_1 \left( g^{\mu\rho} g^{\nu\sigma} - g^{\mu\nu} g^{\rho\sigma} \right) g^{\alpha\beta} \right. \nn
& \left. + \half \xi_2 \left( g^{\alpha\nu} g^{\beta\rho} g^{\sigma\mu} + g^{\alpha\rho} g^{\beta\mu} g^{\sigma\nu} - g^{\alpha\nu} g^{\beta\mu} g^{\sigma\rho} - g^{\alpha\rho} g^{\beta\sigma} g^{\mu\nu} \right) \right] \nabla_\rho \nabla_\sigma \left( A_\alpha A_\beta \right) 	\per
\end{align}
The $F^2$ terms are the familiar stress-energy tensor for the Einstein-Maxwell theory (massless electromagnetism).  The first square brackets are terms arising from the vector field's mass and nonminimal coupling to gravity.  The second square brackets only contains terms from nonminimal gravitational involving derivative terms.  Note that 
\begin{align}
\nabla_\rho\nabla_\sigma \left( A_\alpha A_\beta \right) & = \nabla_\rho \left( \nabla_\sigma A_\alpha \right) A_\beta + \left( \nabla_\rho A_\alpha \right) \left( \nabla_\sigma A_\beta \right) + \left( \nabla_\sigma A_\alpha \right) \left( \nabla_\rho A_\beta \right) + A_\alpha \nabla_\rho \left( \nabla_\sigma A_\beta \right) \per
\end{align}
We calculate the trace to be
\begin{align}\label{eq:RV_T_trace}
\tensor{T}{^\mu_\mu} & = - m^2 g^{\alpha\beta} A_\alpha A_\beta + \left[ 3 \xi_1 g^{\rho\sigma} g^{\alpha\beta} + \half \xi_2 \left( g^{\alpha\sigma} g^{\beta\rho} - g^{\alpha\beta} g^{\rho\sigma} - 3 g^{\alpha\rho} g^{\beta\sigma} \right) \right] \nabla_\rho \nabla_\sigma \left( A_\alpha A_\beta \right) 	\per
\end{align}
Note that $\tensor{T}{^\mu_\mu} = 0$ for a massless ($m = 0$) and minimally-coupled ($\xi_1 = \xi_2 = 0$) vector field.  This calculation reveals that a massless vector with a minimal coupling to gravity is conformally coupled to gravity, so particle production will depend upon $m$ and/or $(\xi_1,\xi_2)$.  Finally, we calculate the energy density,  $\rho = g_{\mu 0} g_{\nu 0} T^{\mu \nu}$ where $g_{\mu\nu} = (1, -a^2, -a^2, -a^2)$.  We write
\begin{align}
\rho & = \rho_1 + \rho_2 + \rho_3 + \rho_4 + \rho_5 + \rho_6 \com
\end{align}
where
\begin{subequations}
\begin{align}
\rho_1 & = \frac{1}{4} \left( g_{00} g^{\alpha\gamma} g^{\beta\delta} - 4 \delta_0^\alpha \delta_0^\gamma g^{\beta \delta} \right) F_{\alpha\beta}  F_{\gamma\delta} \\
\rho_2 & = 	m^2 \left( \delta_0^\alpha \delta_0^\beta - \half g_{00} g^{\alpha\beta} \right) A_\alpha A_\beta \\ 
\rho_3 & = 	- \xi_1 \left( R \delta_0^\alpha \delta_0^\beta + G_{00} g^{\alpha\beta} \right) A_\alpha A_\beta \\ 
\rho_4 & = - \xi_2 \left( \delta_0^\alpha g_{0\nu} R^{\nu\beta} + \delta_0^\beta g_{0\nu} R^{\nu\alpha} - \half g_{00} R^{\alpha\beta} \right) A_\alpha A_\beta \\ 
\rho_5 & = \xi_1 \left( \delta_0^\rho \delta_0^\sigma - g_{00} \, g^{\rho\sigma} \right) g^{\alpha\beta} \nabla_{\rho} \nabla_{\sigma} \left( A_\alpha A_\beta \right) \\ 
\rho_6 & = \half \xi_2 \left( \delta_0^\alpha \delta_0^\sigma g^{\beta\rho} + \delta_0^\beta \delta_0^\sigma g^{\alpha\rho} - \delta_0^\alpha \delta_0^\beta g^{\sigma\rho} - g_{00} \, g^{\alpha\rho} g^{\beta\sigma} \right) \nabla_{\rho} \nabla_{\sigma} \left( A_\alpha A_\beta \right) \per
\end{align}
\end{subequations}

Now we specialize to the FRW geometry.  Just as was done for massive spin-1 fields in Minkowski space, it is convenient to remove the auxiliary field and decompose the vector field into transverse and longitudinal mode functions. 

In component form the action of \pref{eq:RV_action_1} assuming the FRW metric is [cf.\ \eref{eq:actionconponents}]
\begin{align}\label{eq:aactionconponents}
S\left[A_\mu(t,\xvec)\right] = & \int d^4x\left[ \half a \left(\partial_tA_i-\partial_iA_t\right)^2 - \tfrac{1}{4} a^{-1} \left(\partial_iA_j-\partial_jA_i\right)^2  \right. \nn 
& \left. + \half a^3 m^2_{\mathrm{eff},t} A_t^2 - \half a m^2_{\mathrm{eff},x} A_i^2 \right] \com
\end{align}
where we have defined 
\begin{subequations}\label{eq:meff_ts_def}
\begin{align} \label{eq:mefft}
m_{\mathrm{eff},t}^2 & \equiv m^2 - \xi_1 R - \frac{1}{2} \xi_2 R - 3 \xi_2 H^2 \\ \label{eq:meffx}
m_{\mathrm{eff},x}^2 & \equiv m^2 - \xi_1 R - \frac{1}{6} \xi_2 R + \xi_2 H^2 \com
\end{align}
\end{subequations}
which correspond to effective masses for the time-like and space-like components. 
As in Minkowski space, $A_t$ does not have a kinetic term; it is an auxiliary field. 

The field equations and energy density for the FRW metric in component form are [cf.\ \eref{eq:eomcomponents}]
\begin{align}
\left[\delta_{ij}\partial_t^2 + \delta_{ij} H \partial_t - a^{-2} \left(\delta_{ij}\partial_k^2 - \partial_i\partial_j\right) + \delta_{ij}m_{\mathrm{eff},x}^2\right] A_j - \partial_i(\partial_t+H)A_t & = 0 \nn
\left[a^{-2}\partial_j^2 - m_{\mathrm{eff},t}^2\right] A_t - a^{-2}\partial_t\partial_jA_j & = 0 \com
\end{align} 
\begin{align}
\rho & = \biggl[
\half a^{-2} (\partial_0 A_i)^2 + a^{-2} (\partial_0 A_i)(\partial_i A_0) + \half a^{-2} \left( 1 + 4 \xi_1 + 2 \xi_2 \right) (\partial_i A_0)^2 \nn 
& \qquad \quad + \half a^{-4} \left( 1 - 4 \xi_1 \right) (\partial_i A_j)^2 - \half a^{-4} \left( 1 + \xi_2 \right) (\partial_i A_j) (\partial_j A_i) - \half a^{-4} \xi_2 (\partial_i A_i) (\partial_j A_j) \nn
& \qquad \quad 	- a^{-4} \xi_2  A_i \partial_i \partial_j A_j + a^{-2} \left( 2 \xi_1 + \xi_2 \right) A_0 \partial_i^2 A_0 - 2 a^{-4} \xi_1  A_j \partial_i^2 A_j \biggr] \nn 
& \quad + \biggl[ -3( 2\xi_1 H + \xi_2 H) A_0 \partial_0 A_0 + a^{-2} \left( 6 \xi_1 H + \xi_2 H \right) A_i \partial_0 A_i + 2 a^{-2} \xi_2 H A_i \partial_i A_0 \nn
& \qquad \quad	+ 2 a^{-2} \xi_2 H  A_0 \partial_i A_i \biggr\} + \biggl\{ \half \left( m^2 - 2 \xi_1 R - 6 \xi_1 H^2 - \xi_2 R - 12 \xi_2 H^2 \right) A_0^2\nn 
& \qquad \quad + \half a^{-2} \left( m^2 - 6 \xi_1 H^2 - 2 \xi_2 H^2 \right) A_i^2 \biggr] \per
\end{align}
In the expression for $\rho$ we have grouped the terms based on the number of derivatives of the field.  In the limit $m_{\mathrm{eff},x}^2 =m_{\mathrm{eff},t}^2 \to m^2$ and $a\to 1=\mathrm{const.}$, we recover the Minkowski result \pref{eq:rho1mink}.  To gain some intuition it is useful to consider static field configurations and to take $A_0 = 0$, which causes the energy density to reduce to 
\begin{align}\label{eq:rhomv}
\rho & = \half a^{-4} \left( 1 - 4 \xi_1 \right) (\partial_i A_j)^2 - \half a^{-4} \left( 1 + \xi_2 \right) (\partial_i A_j) (\partial_j A_i) - \half a^{-4} \xi_2 (\partial_i A_i) (\partial_j A_j) \nn 
& \quad	 - a^{-4} \xi_2 A_i \partial_i \partial_j A_j - 2 a^{-4} \xi_1 A_j \partial_i^2 A_j + \half a^{-2} \left( m^2 - 6 \xi_1 H^2 - 2 \xi_2 H^2 \right) A_i^2 \per
\end{align}
If $\xi_1=\xi_2=0$, for a relativistic vector field we find $\rho \propto a^{-4}$ from the gradient terms, and once the field becomes non-relativistic we have $\rho \propto a^{-2}$ from the non-gradient terms, which is notably different from the behavior of a non-relativistic scalar field for which $\rho \propto a^0$.  The origin of the difference is that $g^{\mu\nu}$ appears in the mass term for vectors.

Expanding the field in terms of mode functions \pref{eq:vecfour}, the action becomes (again $k^2\equiv \kabs^2$)
\begin{align}\label{eq:RV_action_3}
S[A_\mu(t,\xvec)] = & \int \!dt\, \int \!\! \frac{d^3 \kvec}{(2\pi)^3} \, \biggl[ 
\tfrac{i}{2} a \kvec_i \, A_t^\ast \, (\partial_t A_i) - \tfrac{i}{2} a \kvec_i \, (\partial_t A_i^\ast) \, A_t + \half a \left(k^2 + a^2 m_{\mathrm{eff},t}^2 \right) | A_t |^2 \nn 
& - \tfrac{1}{4} a^{-1} |\kvec_i A_j - \kvec_j A_i|^2  + \half a |\partial_t A_i|^2 - \half a \, m_{\mathrm{eff},x}^2 |A_i|^2 \biggr] \per
\end{align}
Here we have performed the integrals over $\kvec^\prime$ and $\xvec$ to leave only the integral over $\kvec$. Setting $a=1$ and $m_{\mathrm{eff},x}^2 =m_{\mathrm{eff},t}^2 = m^2$ we recover the Minkowski result \pref{eq:actionfourier}.  Again, for notational simplicity we have suppressed the $\kvec$ label on $A_i$ and $A_t$.

In order to solve for the temporal component of the field we rewrite \eref{eq:RV_action_3} as [cf.\ \eref{eq:actionfourier2}]
\begin{align}\label{eq:aactionfourier2}
S\left[A_\mu(t,\xvec)\right] & =  \int \!dt\, \int\ \!\! \frac{d^3\kvec}{(2\pi^3)}\, \left[ \half a\left(k^2 + a^2m_{\mathrm{eff},t}^2\right) \left|A_t + i\dfrac{\kvec_i(\partial_tA_i)}{k^2+a^2m_{\mathrm{eff},t}^2}\right|^2  \right. \nn
& \left. - \half a \frac{\left|\kvec_i(\partial_tA_i)\right|^2}{k^2 + a^2m_{\mathrm{eff},t}^2} - \tfrac{1}{4} a^{-1}\left|\kvec_iA_j-\kvec_j A_i\right|^2 + \half a\left|\partial_tA_i\right|^2 -\half a^2m_{\mathrm{eff},x}^2\left|A_i\right|^2 \right] \per 
\end{align}
Now that $A_t$ is isolated it is clear it is nondynamical and we can solve for it [cf.\ \eref{eq:integrateout}]:
\begin{align}
A_t = - i\dfrac{\kvec_i(\partial_tA_i)}{k^2+a^2m_{\mathrm{eff},t}^2} \per
\end{align}
Then, integrating out $A_t$, the action becomes
\begin{align}
S\left[A_\mu(t,\xvec)\right] & =  \int \!dt\, \int\ \!\! \frac{d^3\kvec}{(2\pi^3)}\,\Biggl[ \half a (\partial_tA_i^*) \left( \delta_{ij}-\dfrac{\kvec_i\kvec_j}{k^2+a^2m_{\mathrm{eff},t}^2}\right) (\partial_tA_j) \nn
& - \half a^{-1}A_i^*\left[(k^2+a^2m_{\mathrm{eff},x}^2)\delta_{ij} -\kvec_i\kvec_j\right]A_j \Biggr] \per
\end{align}
(We have also expanded out the terms in $|\kvec_i A_j - \kvec_j A_i|^2$. )
 
Using again the orthonormal set of basis vectors of \eref{eq:varepsilon} and the mode functions, $A^T_\kvec(x^0)$ and $A^L_\kvec(x^0)$, the action becomes [cf.\ \eref{eq:RV_action_6}]
\begin{subequations} 
\begin{align}\label{eq:aRV_action_6}
S^T & = \sum_{b=1,2} \int dt \, a^{-1} \int \! \! \frac{d^3\kvec}{(2\pi)^3} \, \left[ \half a^2 |\partial_t A_\kvec^{T_b}|^2 - \half \left(k^2 + a^2 m^2_{\mathrm{eff},x}\right) \ |A_\kvec^{T_b}|^2\right] \\ 
S^L  & = \int dt \, a^{-1} \int \! \! \frac{d^3\kvec}{(2\pi)^3} \, \left[ \half \frac{a^2m^2_{\mathrm{eff},t}}{k^2 +a^2m^2_{\mathrm{eff},t}} \, a^2 |\partial_t A_\kvec^L|^2 - \half a^2 m^2_{\mathrm{eff},x} \, |A_\kvec^L|^2 \right] 	\per
\end{align} 
\end{subequations}
Before proceeding further we express $S^T$ and $S^L$ in conformal time $\eta$:
\begin{subequations} \begin{align}
S^T & = \sum_{b=1,2} \int d\eta \, \int \! \! \frac{d^3\kvec}{(2\pi)^3} \, \left[ \half  |\partial_\eta A_\kvec^{T_b}|^2 - \half  \left(k^2 + a^2 m^2_{\mathrm{eff},x}\right) \ |A_\kvec^{T_b}|\right]  \label{eq:steta}\\ 
S^L & = \int d\eta \, \int \! \! \frac{d^3\kvec}{(2\pi)^3} \, \left[ \half  \frac{a^2m^2_{\mathrm{eff},t}}{k^2 +a^2m^2_{\mathrm{eff},t}} \, |\partial_\eta A_\kvec^L|^2 - \half a^2m^2_{\mathrm{eff},x} \, |A_\kvec^L|^2 \right] \label{eq:sleta}	\per
\end{align} \end{subequations}
Here we see that the action for the individual transverse modes is precisely the action for a scalar field with $m^2_{\mathrm{eff},x}$ defined in \eref{eq:meffx} (recall for scalars $\meff^2=m^2+(1-6\xi)R/6$).  We also see that we have to have a field redefinition to have a proper action for the longitudinal action.  We also note that if we keep nonminimal terms in the action $m^2_{\mathrm{eff},t}$ can be negative and the kinetic term could be negative, leading to a ghost-like action  \cite{Nakayama:2019rhg}; we will consider only minimal gravitational interactions.  As to the rationale for only considering minimal gravitational interactions, we note that the addition of the nonminimal terms in \eref{eq:RV_action_1} breaks gauge symmetry (as does the mass term, but that might arise from the Stuckelberg trick).

To have a correct kinetic term for the longitudinal mode we define $\chi_\kvec^L$ as [cf.\ \eref{eq:alphil}]
\begin{align} 
A^L_\kvec(\eta) = \kappa_k(\eta) \chi^L_\kvec(\eta) \quad \mathrm{with} \quad \kappa_k(\eta) =  \sqrt{\frac{k^2+a^2m^2_{\mathrm{eff},t}}{a^2m^2_{\mathrm{eff},t}}} \  \per
\end{align}
To simplify notation we will drop the superscript $L$ and the subscript $\kvec$ on $\chi^L_\kvec$ and suppress the $k$ subscript on $\kappa_k$, with the understanding that $\chi$ represents the Fourier mode for the longitudinal component, and that $\kappa$ (not to be confused with $\kappa=8\pi G$) is a function of $k$ and $\eta$.  For future use we note
\begin{align}\label{eq:alphak}
\frac{\partial_\eta\kappa}{\kappa} = -\frac{k^2aH}{k^2+a^2m^2_{\mathrm{eff},t}} \left(1+\frac{1}{aH}\frac{\partial_\eta m_{\mathrm{eff},t}}{m_{\mathrm{eff},t}} \right) \per
\end{align}
With the field redefinition \pref{eq:alphil} the kinetic term is 
\begin{align}
\Lcal_\mathrm{kinetic} & = \half |\partial_\eta\chi|^2 + \half \left(\frac{\partial_\eta\kappa}{\kappa}\right)^2|\chi|^2 + \half \frac{\partial_\eta\kappa}{\kappa}\left[ (\partial_\eta\chi)\chi + \chi (\partial_\eta\chi) \right]  \nn
& =  \half |\partial_\eta\chi|^2 + \half\left[\left(\frac{\partial_\eta\kappa}{\kappa}\right)^2  - \half\partial_\eta\left(\frac{\partial_\eta\kappa}{\kappa}\right)  \right] |\chi|^2 \com
\end{align}
where the second equality is the result of an integration by parts. This leads to an action for the longitudinal component of
\begin{align}\label{eq:phil}	
S^L & = \int d\eta \, \int \! \! \frac{d^3\kvec}{(2\pi)^3} \,   \left\{ \half |\partial_\eta \chi|^2 - \half \left[ k^2\dfrac{m^2_{\mathrm{eff},x}}{m^2_{\mathrm{eff},t}} + a^2m^2_{\mathrm{eff},x} + \partial_\eta\left(\frac{\partial_\eta\kappa}{\kappa}\right) -  \left(\dfrac{\partial_\eta\kappa}{\kappa}\right)^2 \right] |\chi|^2 \right\} \per
\end{align}

To summarize, the transverse and longitudinal components are independent, with actions
\begin{subequations}
\begin{align}
S^T & = \sum_{b=1,2} \int d\eta \, \int \! \! \frac{d^3\kvec}{(2\pi)^3} \, \left( \half  |\partial_\eta A_\kvec^{T_b}|^2 - \half \omega_T^2 |A^{T_b}_\kvec|^2 \right)  \label{eq:transaction } \\
S^L & = \int d\eta \, \int \! \! \frac{d^3\kvec}{(2\pi)^3} \, \left( \half |\partial_\eta \chi|^2 - \half \omega_L^2 |\chi|^2 \right) \com
\end{align}
\end{subequations}
where we have defined the squared natural frequencies to be
\begin{subequations}\label{eq:omega_TL_def}
\begin{align}
\omega_T^2(\eta) & \equiv k^2 + a^2 m_{\mathrm{eff},x}^2 \\ 
\omega_L^2(\eta) & \equiv k^2 \frac{m_{\mathrm{eff},x}^2}{m_{\mathrm{eff},t}^2} + a^2 m_{\mathrm{eff},x}^2 + \partial_\eta\left(\frac{\partial_\eta\kappa}{\kappa}\right) -  \left(\dfrac{\partial_\eta\kappa}{\kappa}\right)^2 \per
\end{align}
\end{subequations}
The mode functions $A_\kvec^{T_b}$ and $\phi_\kvec^L$ satisfy the mode equations
\begin{subequations}\label{eq:modes}
\begin{align}
\partial_\eta^2 A_\kvec^{T_b} + \omega_T^2 A_\kvec^{T_b} & = 0 \\
\partial_\eta^2 \chi + \omega_L^2 \chi & = 0 \per \label{eq:chipp}
\end{align}
\end{subequations} 
The frequencies in general are rather complicated, but they simplify if we consider $\xi_1=\xi_2=0$.  With that choice $m_{\mathrm{eff},x}^2=m_{\mathrm{eff},t}^2=m^2$, where $m^2$ is a constant, leading to  
\begin{subequations}
\begin{align}
\omega_T^2(\eta) & \equiv k^2 + a^2 m^2 \\ 
\omega_L^2(\eta) & \equiv k^2 + a^2 m^2 + \frac{1}{6}\frac{k^2}{k^2+a^2m^2} a^2R +3\frac{k^2}{(k^2+a^2m^2)^2}a^2H^2a^2m^2 \label{eq:omegal2}\per
\end{align}
\end{subequations}
Thus, the transverse mode behaves as a conformally-coupled scalar field ($\xi=1/6$) with two degrees of freedom.  In the limit $am \ll k$, $\omega_T^2=k^2$ is time-independent and the mode \textit{will not} be populated by expansion.  That is not true for the longitudinal mode.  In the limit $am\ll k$, $\omega_L^2(\eta) = k^2+a^2R/6$, and the longitudinal component appears as a massless, minimally-coupled scalar field ($\xi=0$), which \textit{will} be populated in expansion.   In the late-time limit $k\ll am$, $R\ll m^2$, and $H^2\ll m^2$ the frequency of both modes have the expected form, $\omega^2=a^2m^2$.

%------------------------------ 
\section{Gravitational particle production (GPP) during inflation \label{sec:GPP}}
%------------------------------ 

Now we turn to the phenomenon of gravitational particle production.  The idea that the expansion of the universe may result in particle production goes back at least as far as a 1939 paper by Erwin Schr\"{o}dinger \cite{Schrodinger:1939:PVE}. Its modern field-theory incarnation started with the early work of Parker (see, e.g. \cite{Parkerthesis}).  Quantum field theory in curved spacetime has been well developed (see e.g., \cite{BirrellDavies:1982}), and in the context of inflation it has been studied with an eye towards producing dark matter; first studied assuming the spectator field was a fermion or scalar \cite{Chung:1998bt, Kuzmin:1998uv}, and more recently assuming the spectator field is a massive vector \cite{Graham:2015rva}. In this paper we focus on the massive vector case. 

The basic idea behind GPP is that unless the terms in the Lagrangian involving the field are invariant under conformal (Weyl) transformations (operationally this means that the trace of the stress-energy tensor must not vanish) a rapid expansion of the universe will ``pull'' particles from the vacuum to propagate as real particles. 

It is convenient to calculate GPP by calculating the Bogoliubov coefficient relating the early-time and late-time vacua.  In a system with a time-dependent Hamiltonian the late-time creation and annihilation operators are related to the early-time ones by 
\begin{align}
\hat{a}_\kvec^\mathrm{late} & = \alpha^*_{-\kvec} \hat{a}_k^\mathrm{early} - \beta_{-\kvec}\hat{a}_{-\kvec}^{\dagger\mathrm{early}} \nn
\hat{a}_\kvec^{\dagger\mathrm{late}} & =  \alpha_{-\kvec}\hat{a}_{-\kvec}^{\dagger\mathrm{early}} - \beta^*_{-\kvec} \hat{a}_\kvec^\mathrm{early}  \per
\end{align}
The early-time observer defines a vacuum by 
\begin{align}\label{eq:earlyvacuum}
\hat{a}_\kvec^\mathrm{early} \ket{0^\mathrm{early}} = 0  \ket{0^\mathrm{early}} \  \forall \ \kvec \com
\end{align}
which implies that the late-time observer detects particles:
\begin{align}
\langle \hat{N}^\mathrm{late}\rangle = \int \frac{d^3\kvec}{(2\pi)^3} \ \expval{0^\mathrm{early}}{\hat{a}_\kvec^{\dagger\mathrm{late}}\,\hat{a}_\kvec^\mathrm{late}}{0^\mathrm{early}} = V\int \frac{dk}{k} \frac{k^3}{2\pi^2}\, \left|\beta_k\right|^2 \per
\end{align}

The mode equations \pref{eq:modes} are solved subject to initial conditions \pref{eq:earlyvacuum} to obtain $\chi_k(\eta)$.  The modulus of the second Bogoliubov coefficient is extracted from the solution to the mode equations:
\begin{align}
\left|\beta_k\right|^2 & = \lim_{\eta\to\infty} \left[ \frac{\omega_k}{2} \left|\phi_k\right|^2 + \frac{1}{2\omega_k} \left|\partial_\eta\phi_k\right|^2 + \frac{i}{2}\left(\phi_k\partial_\eta\phi^*_k - \phi^*_k\partial_\eta\phi_k \right) \right] \nn
& = \lim_{\eta\to\infty} \left[ \frac{\omega_k}{2} \left|\phi_k\right|^2 + \frac{1}{2\omega_k} \left|\partial_\eta\phi_k\right|^2 - \frac{1}{2} \right]
 \com
\end{align}
where $\phi_k$ stands for either $A_\kvec^{T_b}$ or $\chi$ and $\omega_k$ is the corresponding value of $\omega_T$ or $\omega_L$.  The factor of $\phi_k\partial_\eta\phi^*_k - \phi^*_k\partial_\eta\phi_k =i$ as demanded by the  commutation relations.  We will define the spectrum of the mode function, $n_k$, and the comoving number density of particles, $na^3$, as
\begin{align}
n_k = \frac{k^3}{2\pi^2} \left|\beta_k\right|^2 \nn
na^3 = \int \frac{dk}{k}\ n_k
\end{align}

Now for initial conditions.  For GPP in the inflationary era the early-time limit ($\eta\to-\infty$) corresponds to $a\to0$ and $a^2R\to0$, which implies $\omega_k^2 \to k^2$.  As $\eta\to-\infty$ the modes are deep within the Hubble radius and their mode equation is approximately that of Minkowski space.  Thus, the natural initial condition, the so-called Bunch-Davies initial condition, on the mode functions $\phi=\{A_\kvec^{T_b},\chi\}$ is
\begin{align}
\lim_{\eta\to-\infty} \phi_k(\eta) = \frac{1}{\sqrt{2k}}\ e^{-ik\eta} \per
\end{align}
(The factor of $1/\sqrt{2k}$ ensures the commutation relations are properly normalized.)

Gravitational particle production of the transverse mode is exactly the same as the well-studied case of GPP of conformally-coupled scalars.  Conformal symmetry is exact in the limit $m\to0$, so the result for the transverse component must vanish as $m\to0$.  That is not true for the longitudinal mode.

%============
\begin{figure}[p]
\begin{center}
\includegraphics[width=0.99\textwidth]{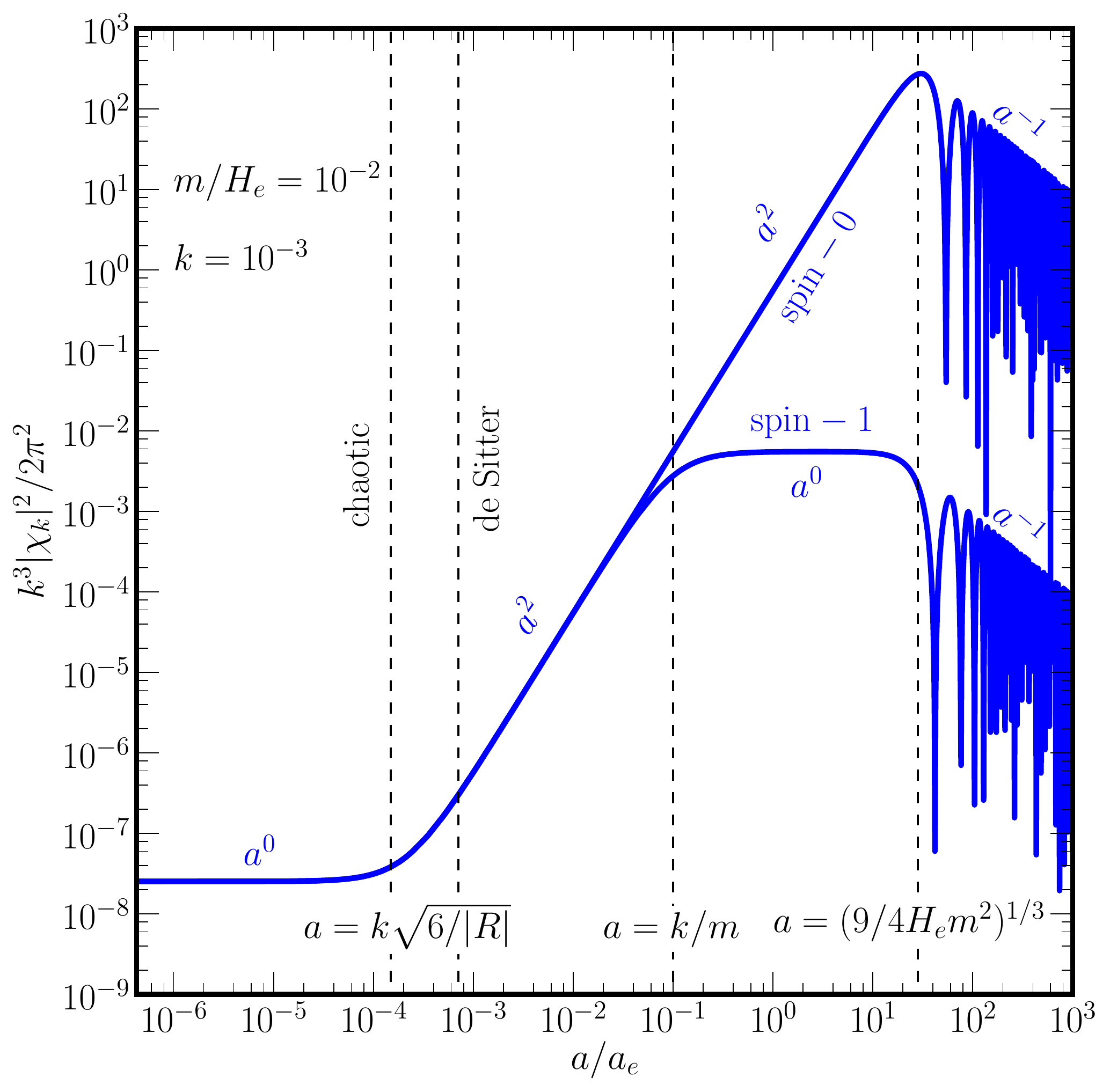}\\
\caption{Numerical results for the evolution of $k^3|\chi_k|^2/2\pi^2$ for $m/H_e=10^{-2}$ and $k=10^{-4}$.  Shown are the results for the longitudinal component of a massive vector field (``spin--1'') and for comparison the evolution of a massive minimally-coupled scalar field (``spin-0''), which corresponds to the transverse component.  The left-most pair of vertical dashed lines corresponds approximately to the time when this mode left the horizon during inflation (assuming either chaotic $m^2 \phi^2$ inflaton, or simply de Sitter).  The dashed line labeled $a=k/m$ corresponds to the time when this mode became nonrelativistic.  Inflation ends at $a/a_e = 1$.  The right-most dashed line corresponds to the time when $H(a) \approx m$. 
\label{fig:compscalarvector}}  
\end{center}
\end{figure}
%============

An example of the evolution in $a$ of $|\chi_k|^2$ for the longitudinal component of a vector field and for a minimally-coupled scalar (which is identical to the transverse component of the vector field) is illustrated in \fref{fig:compscalarvector} for a particular choice of $m/H_e=10^{-2}$ and $k=10^{-3}$. Note the region starting at $a=k/m$ where $|\chi|^2$ is constant for the longitudinal component of a vector, but grows as $a^2$ for a minimally-coupled scalar.

%============
\begin{figure}[t]
\begin{center}
\includegraphics[width=0.495\textwidth]{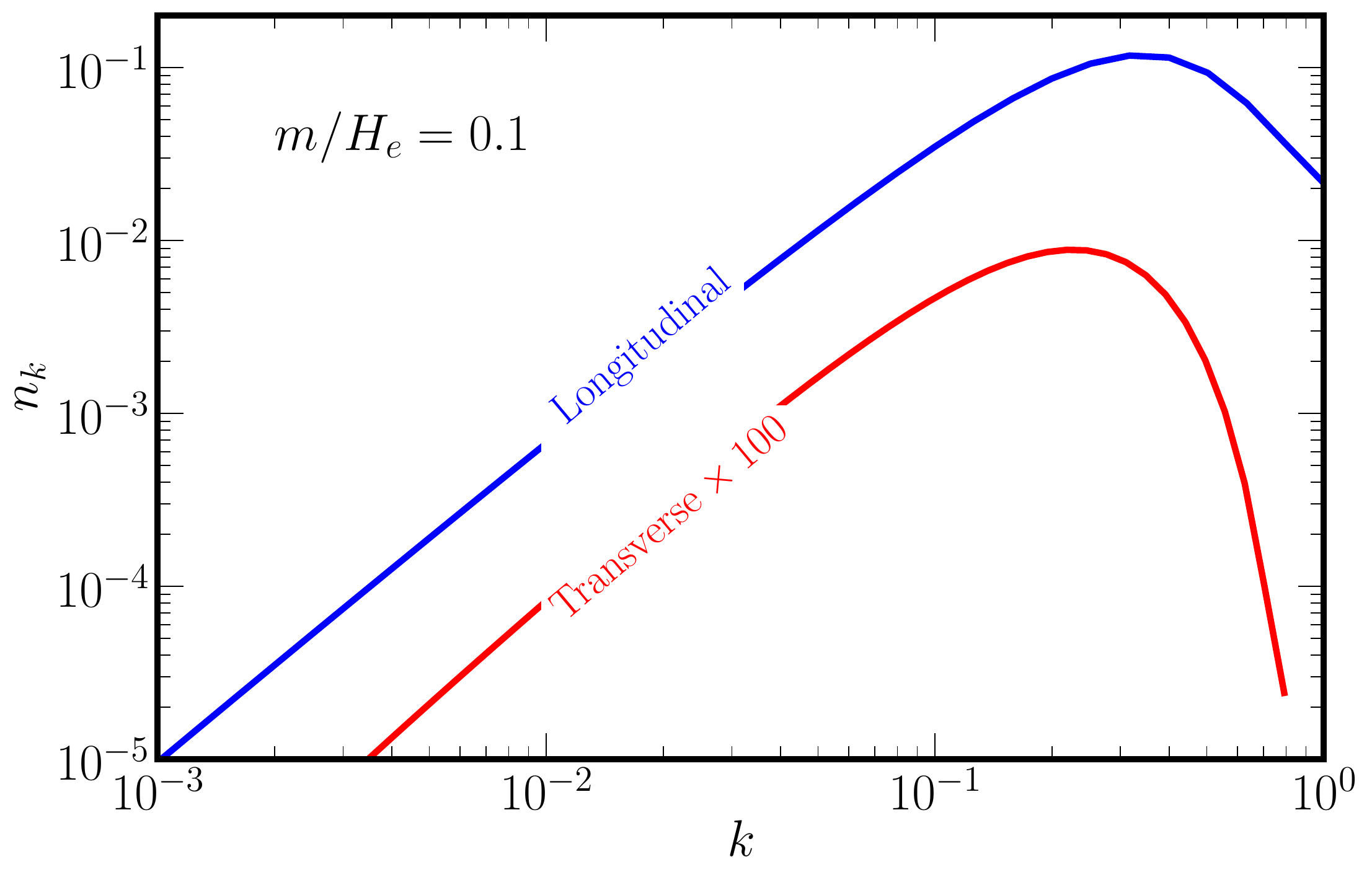}
\includegraphics[width=0.495\textwidth]{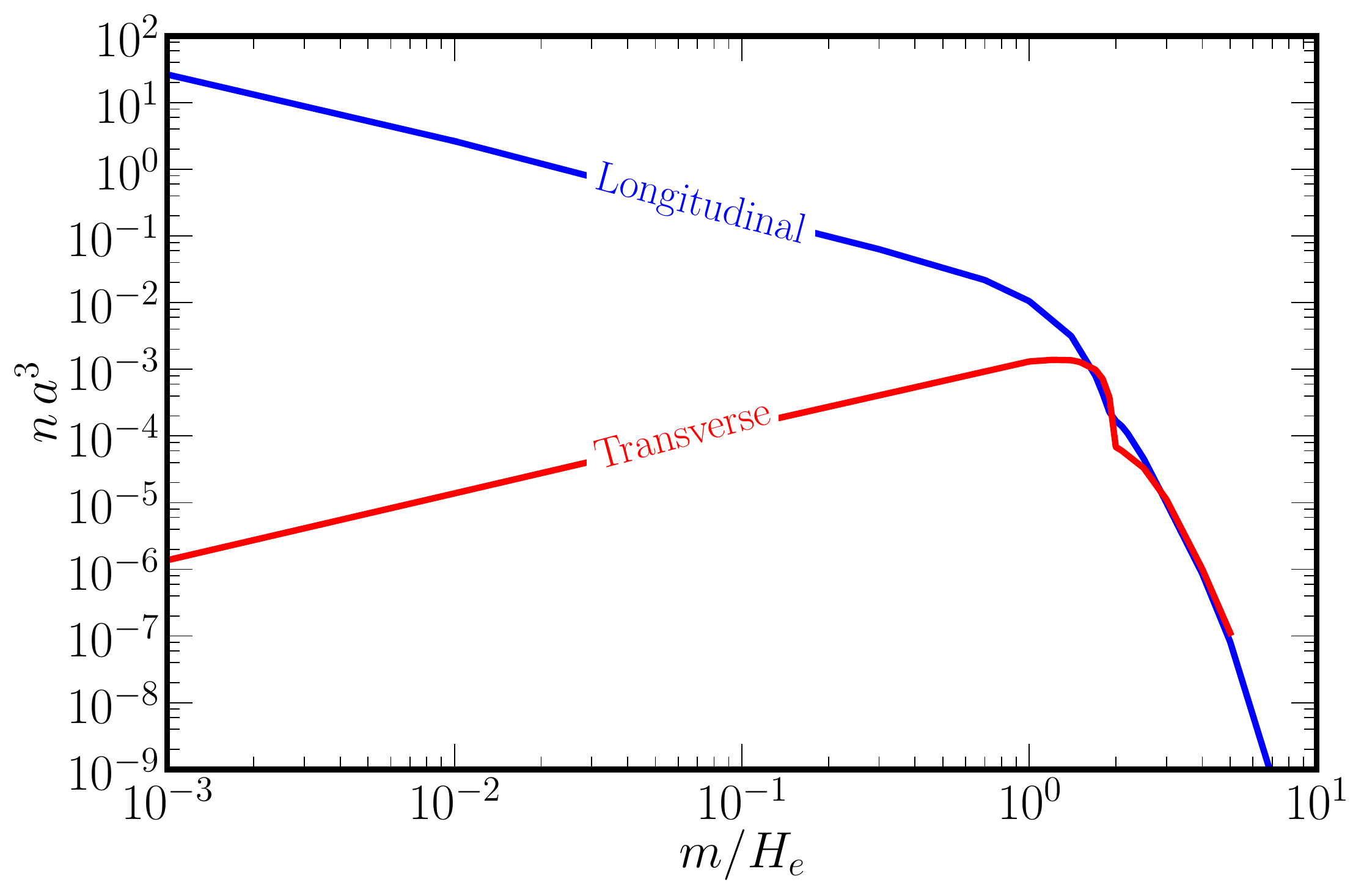}  
\caption{\label{fig:vector_spectrum} Left panel: Numerical results for the spectrum of gravitationally-produced particles in the chaotic model for massive vector fields assuming $m/H_e=0.1$. (Note that the transverse mode has been multiplied by a factor of $10^2$.) Recall that $k = k/a_e H_e$, $n_k = n_k / a_e^3 H_e^3$, $n \, a^3 = n \, a^3 / a_e^3 H_e^3$, and we set $a_e H_e = 1$.}
\end{center}
\end{figure}
%============

A numerical solution for the massive-vector spectrum with $m/H_e=0.1$ is shown in \fref{fig:vector_spectrum}.  Notice that although the frequency of the longitudinal mode resembles a minimally-coupled scalar as modes cross the Hubble radius during inflation, the spectrum at small $k$ does not resemble the scalar spectrum, which grows at small $k$.  The integration of the mode functions is also shown in \fref{fig:vector_spectrum}.   For this figure we assumed that the mode evolves to become nonrelativistic ($k/a<m$) with $H/a<m$.  In that region $|\chi_k|^2$  oscillates and decays as $a^{-1}$, i.e., it behaves as nonrelativistic matter.

From the numerical results we see the expected result that for $m>H_e$ the mode function is exponentially damped as $e^{-\pi m/H_e},$\footnote{This occurs in the chaotic and analytic models we will consider.  It is possible to evade this suppression (at least for a while) in other models of inflation like hilltop inflation \cite{Ema:2018ucl,Chung:2018ayg}.} and that the modes are exponentially damped for $k>1$.  

%------------------------------ 
\section{Analytic approximation to the comoving number density \label{sec:density}}
%------------------------------ 

The goal in this section is to obtain an analytic approximation for $na^3$ for the longitudinal component where we consider the possibility that reheating to a radiation-dominated phase occurs before the modes have reached the point where a particle description is appropriate.

In terms of dimensionless quantities $\alpha$, $\mu$, $h$, the frequency for the longitudinal component, \eref{eq:omegal2}, becomes
\begin{align}
\omega_L^2=k^2 + \alpha^2\mu^2 + \frac{k^2}{k^2+\alpha^2\mu^2}\alpha^2\frac{R}{6H_e^2} + 3 \frac{k^2}{(k^2+\alpha^2\mu^2)^2}\alpha^4h^2\mu^2 \per
\end{align}

In dS, $h=1$, $R/6H_e^2=-2$, and assuming $\mu<1$ there are four possible dominant terms in  $\omega_L^2$ in four regions of $\alpha$ and $\mu$, denoted by $\IdS$--$\IVdS$:
\begin{align}
\begin{array}{ll}
\omega_L^2 = k^2+\alpha^2\mu^2 - 2\dfrac{k^2}{k^2+\alpha^2\mu^2}\alpha^2 + 3\dfrac{k^2}{(k^2+\alpha^2\mu^2)^2}\alpha^4\mu^2 & \quad\quad \mathrm{dS} \quad 0\leq\alpha<1 \\[3ex]
\phantom{\omega_L^2} = \left\{ 
\begin{array}{cll} 
	k^2        & 1>k>\sqrt{2}\alpha & \quad\quad \IdS \\[2ex]
	-2\alpha^2 & \sqrt{2}\alpha>k>\alpha\mu  & \quad\quad \IIdS \\[2ex]
	\dfrac{k^2}{\mu^2} & \alpha\mu>k>\alpha\mu^2 & \quad\quad \IIIdS\\[2ex]
	\alpha^2\mu^2 & \alpha\mu^2>k>0 & \quad\quad \IVdS \per  
\end{array} \right.
\end{array}
\end{align}

%============
\begin{figure}[t]
\begin{center}
\includegraphics[width=.95\textwidth]{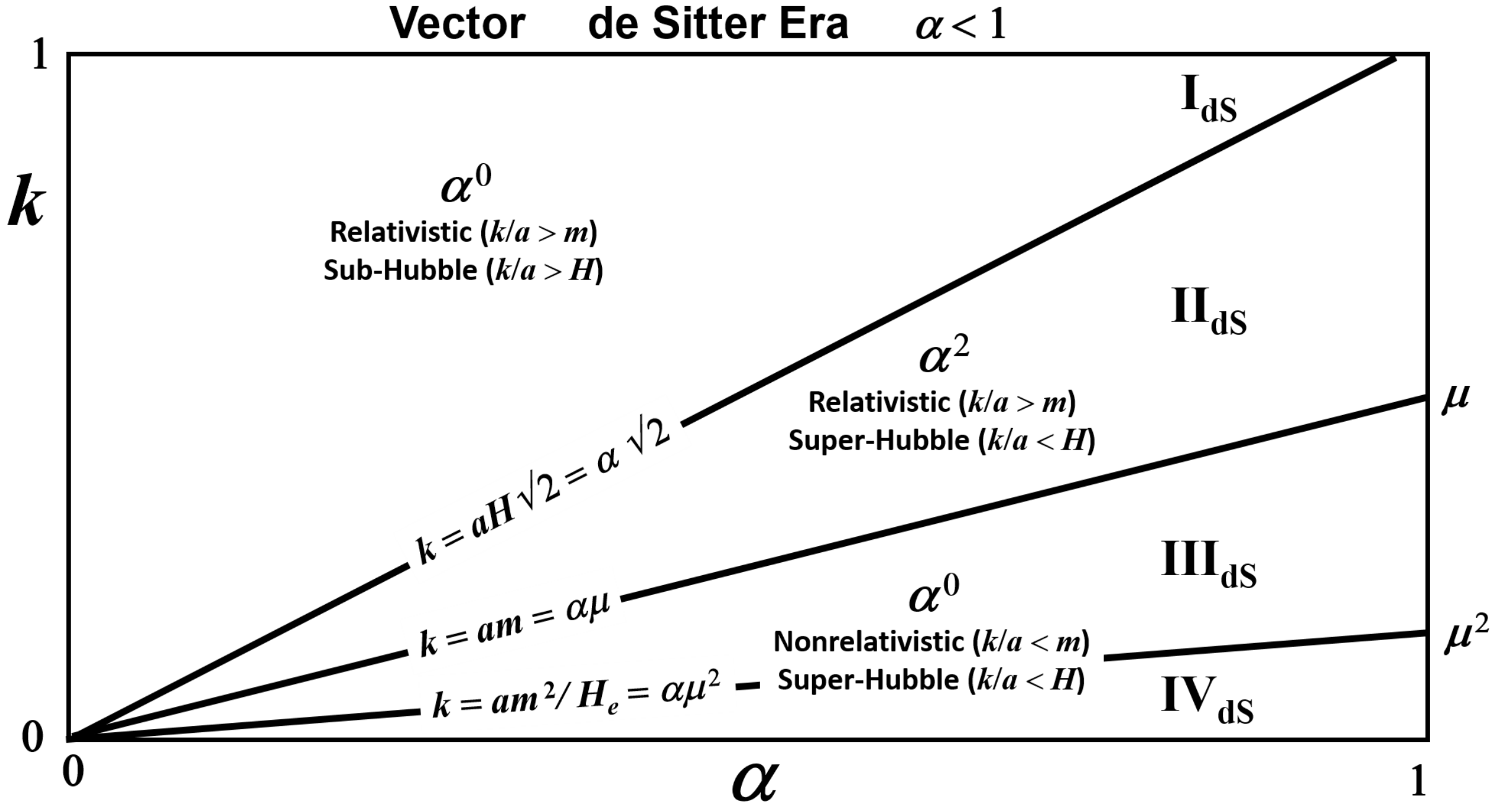}\\
\caption{Regions in the $k$--$\alpha$ plane for evolution in the dS phases.  In different regions we indicate the scaling of $|\chi_k|^2$. The line denoted $k=\alpha\sqrt{2}$ is the line for $k=\alpha\sqrt{|R|/6H_e^2}=aH/\sqrt{2}$; above the line the mode is sub-Hubble and below it is super-Hubble. The line $k=\alpha\mu$ is the line that devides the relativistic ($k>am$) and nonrelativistic ($k<am$) regions.  \label{fig:dS_vector}}
\end{center}
\end{figure}
%============

In MD, $h=\alpha^{-3/2}$, and $R/6H_e^2=-\half \alpha^{-3}$, and again there are four possible dominant terms in $\omega_L^2$; they are in regions $\IMD$--$\IVMD$:\footnote{When considering transitions between different regions in the MD era we will be cavalier about numerical factors of order unity. We will use ``$\simeq$'' to indicate equations where  we have dropped order unity numerical factors.}
\begin{align}
\begin{array}{ll}
\omega_L^2 = k^2+\alpha^2\mu^2 - \dfrac{1}{2}\dfrac{k^2}{k^2+\alpha^2\mu^2}\alpha^{-1} + 3\dfrac{k^2}{(k^2+\alpha^2\mu^2)^2} \mu^2\alpha & \quad\quad \mathrm{MD} \quad 1<\alpha<\alphaRH \\[3ex]
\phantom{\omega_L^2} = \left\{
\begin{array}{cll} 
	k^2 & 1>k>\mathrm{Max}(\alpha^{-1/2},\ \alpha\mu) & \quad\quad  \IMD \\[2ex]
	-\dfrac{1}{2}\dfrac{1}{\alpha} & \alpha^{-1/2}>k\gtrsim\alpha\mu & \quad\quad \IIMD \\[2ex]
	\dfrac{5}{2}\dfrac{k^2}{\mu^2}\dfrac{1}{\alpha^3} & \alpha\mu>k\gtrsim\alpha^{5/2}\mu^2 & \quad\quad \IIIMD \\[2ex]
		\alpha^2\mu^2 & \mathrm{Min}(\alpha^{5/2}\mu^2,\ \alpha\mu) \gtrsim k>0 & \quad\quad \IVMD \per
\end{array} \right.  
\end{array}
\end{align}

In RD, $h=\alphaRH^{1/2}/\alpha^2$, $R/6H_e^2=0$, and the dominant terms in $\omega_L^2$ are\footnote{Since $R=0$ in RD, there is no region corresponding to $\IIMD$.}
\begin{align}
\begin{array}{ll}
\omega_L^2 = k^2+\alpha^2\mu^2 + 3\dfrac{k^2}{(k^2+\alpha^2\mu^2)^2} \alphaRH\mu^2 & \quad \mathrm{RD} \quad \alphaRH<\alpha<\infty \\[3ex]
\phantom{\omega_L^2} = \left\{
\begin{array}{cll} 
	k^2+3\dfrac{\alphaRH\mu^2}{k^2} & 1>k>\alpha\mu & \quad\quad  \IRD \\[2ex]
	3\dfrac{k^2}{\mu^2}\dfrac{1}{\alpha^4} &  \alpha\mu>k \gtrsim \alpha^3 \mu^2 \alphaRH^{-1/2} & \quad\quad \IIIRD\\[2ex]	
	\alpha^2\mu^2 & \mathrm{Min}(\alpha^3\mu^2\alphaRH^{-1/2},\ \alpha\mu) > k > 0 & \quad\quad \IVRD 
 \per
\end{array} \right.  
\end{array}
\end{align}

The wave equation \pref{eq:chipp} is $\chi_k''(\eta) + \omega_L^2(\eta)\chi_k(\eta) = 0$. We are interested in the scaling of $|\chi_k|^2$ with $\alpha$ when various terms dominate $\omega_L^2$.  In order to solve the wave equation for $|\chi_k|^2$ in various regions, we have to convert the $\alpha$-dependence of $\omega_L^2$ to a $\eta$-dependence using \tref{table:aetaR}.  The wave equations for the various regions are given in \tref{table:solutions}. Let's take each region in turn: 

\begin{enumerate}

\item $\IdS$:  The wave equation and solution in this relativistic sub-Hubble region is 
\begin{align}
\chi_k''+k^2\chi_k  = 0 & \Longrightarrow
\chi_k  = c_1e^{-ik\eta} + c_2e^{ik\eta} \nn & \Longrightarrow |\chi_k|^2 \propto \alpha^0 \com
\end{align}
where we have used the fact that the Bunch-Davies boundary condition yields $c_2=0$.  

\item $\IIdS$: The wave equation and solution in this relativistic super-Hubble region is
\begin{align}
\chi_k''-\dfrac{2}{\eta^2}\chi_k  = 0 & \Longrightarrow
\chi_k = c_1\eta^{-1}+c_2\eta^2 \nn & \Longrightarrow |\chi_k|^2 \propto \alpha^2 \com
\end{align}
where we have only kept the growing mode ($\eta^{-1}=\alpha$).  Together the $k^2$ and $\alpha^2R$ terms give a mode equation that resembles the Mukhanov-Sasaki evolution equation for curvature perturbations \cite{Mukhanov:1988jd,Sasaki:1983kd,Kodama:1985bj}.

\item $\IIIdS$: In this nonrelativistic super Hubble radius region
\begin{align}
\chi_k''+\dfrac{k^2}{\mu^2}\chi_k  = 0 & \Longrightarrow
\chi_k  = c_1e^{-i(k/\mu)\eta} + c_2e^{i(k/\mu)\eta} \nn & \Longrightarrow |\chi_k|^2 \propto \alpha^0 \per
\end{align}
Since $k<\alpha\mu$ in this region implies $k\eta<\mu$, and the magnitude of the argument of the exponentials is small and expansion is justified.  Here we have kept the growing mode. 

\item $\IVdS$:  In this final de Sitter region (also nonrelativistic super Hubble)
\begin{align}
\chi_k''+\dfrac{\mu^2}{\eta^2}\chi_k  = 0 & \Longrightarrow \chi=c_1\eta^{\left(1-\sqrt{1-4\mu^2}\right)/2} + c_2\eta^{\left(1+\sqrt{1-4\mu^2}\right)/2} \nn & \Longrightarrow |\chi_k|^2\propto \alpha^0 \com
\end{align}
where we have only taken the growing mode and have used $\mu<1$.   This region is also relativistic super-Hubble.

\end{enumerate}

The results for dS are also summarized in \tref{table:solutions}.  Various regions and the scaling with $\alpha$ for dS are indicated in \fref{fig:dS_vector}. Also indicated in the figure is the physical significance of various regions: relativistic for $k/a>m$, nonrelativistic for $k/a<m$, super-Hubble-radius for $k/a<H$, and sub-Hubble radius for $k/a>H$. In dS, in the relativistic-sub-Hubble region $|\chi_k|^2\propto \alpha^0$, in the relativistic super-Hubble region $|\chi_k|^2\propto \alpha^2$, and in the nonrelativistic region $|\chi_k|^2\propto \alpha^0$.  For all regions we are assuming $m<H$.

%===================
\begin{table}[p]
\begin{center}
\caption{\label{table:solutions} Relevant solutions to the wave equation assuming a single term in $\omega_L^2$ dominates. \label{table:regions}} 
\begin{tabular}{|l|c|c|c|}\hline
epoch/$k$-range & $\omega_k^2$ & mode equation & $|\chi_k|^2$ $\alpha$-dependence \\ \hline	
de Sitter & & & \\[1ex]
$1>k>\sqrt{2}\alpha$ & $k^2$ & $\chi_k''+k^2\chi_k=0$ & $\alpha^0$  \\ [2ex] 
$\sqrt{2}\alpha>k>\alpha\mu$ & $-2\alpha^2$ & $\chi_k'' - \dfrac{2}{\eta^2}\chi_k=0$ & $\alpha^2$ \\ [2ex]
$\alpha\mu>k>\alpha\mu^2$ & $\dfrac{k^2}{\mu^2}$ & $\chi_k''+\dfrac{k^2}{\mu^2}\chi_k=0$ & $\alpha^0$  \\ [2ex]
$\alpha\mu^2>k>0$ & $\alpha^2\mu^2$ & $\chi_k'' + \dfrac{\mu^2}{\eta^2}\chi_k=0$ & $\alpha^0$ \\[2ex] \hline 
Matter Dominated &&& \\[1ex]
$1>k>\mathrm{Max}(\alpha^{-1/2},\alpha\mu$) & $k^2$ & $\chi_k''+k^2\chi_k=0$ & $\alpha^0$ \\ [2ex]
$\alpha^{-1/2}>k>\alpha\mu$ & $-\dfrac{1}{2}\dfrac{1}{\alpha}$  & $\chi_k'' - \dfrac{2}{\eta^2}\chi_k=0$ & $\alpha^2$ \\[2ex]
$\alpha\mu>k>\alpha^{5/2}\mu^2$ & $\dfrac{5k^2}{2\mu^2}\dfrac{1}{\alpha^3}$ & $\chi_k'' + \dfrac{160k^2}{\mu^2}\dfrac{1}{\eta^6}\chi_k=0$ & $\alpha^0$ \\[2ex]
$\mathrm{Min}(\alpha^{5/2}\mu^2,\alpha\mu)>k>0$ & $\alpha^2\mu^2$ & $\chi_k''+\dfrac{\mu^2}{16}\eta^4\chi_k=0$ & $\begin{array}{ll}\alpha^0 & \quad (\alpha \lesssim \mu^{-2/3}) \\ \alpha^{-1} & \quad (\alpha \gtrsim \mu^{-2/3}) \end{array}$ \\ [2ex] \hline
Radiation Dominated &&&\\[1ex]
$1>k>\alpha\mu$ & $k^2$ & $\chi_k''+\left(k^2+\dfrac{3\alphaRH\mu^2}{k^2}\right)\chi_k=0$ & $\begin{array}{ll} \alpha^2 & (\alpha<\alphaRH^{1/2}k^{-1}) \\[1ex] 
\alpha^0 & (\alpha>\alphaRH^{1/2}k^{-1}) \end{array}$\\ [4ex]
$\alpha\mu>k>\alpha^3\mu^2\alphaRH^{-1/2}$ & $3\dfrac{k^2}{\mu^2}\dfrac{1}{\alpha^4}$ & $\chi_k''+ \dfrac{3k^2}{\mu^2\alphaRH^2}\dfrac{1}{\eta^4}\chi_k=0$& $\alpha^2$ \\[2ex] 
$\mathrm{Min}(\alpha^3\mu^2\alphaRH^{-1/2},\alpha\mu)>k>0$ & $\alpha^2\mu^2$ & $\chi_k''+\alphaRH\mu^2\eta^2\chi_k=0$ & $\begin{array}{ll}\alpha^0 &  (\alpha \lesssim \alphaRH^{1/4}\mu^{-1/2}) \\[1ex] \alpha^{-1} & (\alpha \gtrsim \alphaRH^{1/4}\mu^{-1/2}) \end{array}$ \\ [3ex]\hline
\end{tabular}
\end{center}
\end{table}
%===================

\begin{enumerate}\setcounter{enumi}{4}

\item $\IMD$:  In this relativistic sub-Hubble region the wave equation and solution is 
\begin{align}
\chi_k''+k^2\chi_k  = 0 & \Longrightarrow \chi_k = c_1e^{ik\eta}+c_2e^{-ik\eta} \nn &\Longrightarrow |\chi_k|^2 \propto \alpha^0 \per
\end{align}
Here, some explanation is required.  Since $k>\alpha^{-1/2}$ in this region, the argument of the trigonometric functions, $k\eta \simeq 2k\sqrt{\alpha}$, is much larger than unity, and $\chi_k$ will oscillate with constant amplitude, hence $|\chi_k|^2\propto\alpha^0$.

\item $\IIMD$: The wave equation and solution in $\IIMD$ (relativistic super-Hubble) is
\begin{align}
\chi_k''-\dfrac{2}{\eta^2}\chi_k  = 0 & \Longrightarrow \chi_k = c_1\eta^{-1}+c_2\eta^2 \nn & \Longrightarrow |\chi_k|^2 \propto \alpha^2 \com
\end{align}
where we have only kept the growing mode ($\eta^2=\alpha$ in MD).

\item $\IIIMD$: For this penultimate region (nonrelativistic $H>m$) in MD, the wave equation and solution is
\begin{align}
\chi_k'' + \frac{160k^2}{\mu^2}\frac{1}{\eta^6}\chi_k=0 & \Longrightarrow \chi_k = c_1 x^{-1/4}J_{-1/4}(x) + c_2 x^{-1/4}J_{1/4}(x) \nn & \Longrightarrow |\chi_k|^2 \propto \alpha^0 \com
\end{align}
where here $x = \sqrt{5/2}\,(k/\alpha\mu)$.  In this region $k<\alpha\mu$, and the expansion of the solution for small $x$ yields $\chi_k = c_1\sqrt{2/x} + c_2$.  The mode enters this region at the end of inflation with $\chi_k \propto \alpha^0$.  This boundary condition implies $c_1=0$ and $|\chi_k|^2\propto \alpha^0$. 

\item $\IVMD$: Here, in the final region (nonrelativistic $H<m$) the scaling of the solution to the wave equation with $\alpha$ depends upon whether $\alpha$ is larger or smaller than $\mu^{-2/3}$.  The wave equation and general solution is 
\begin{align}\label{eq:chikinIIIMD}
\chi_k''+\frac{\mu^2}{16}\eta^4\chi_k  = 0 & \Longrightarrow \chi_k = c_1x^{1/6}J_{-1/6}(x) + c_2x^{1/6}J_{+1/6}(x) \nn
& \Longrightarrow |\chi_k|^2 \propto \left\{ \begin{array}{ll} \alpha^0 & \alpha<\mu^{-2/3} \\ \alpha^{-1} & \alpha> \mu^{2/3} \com \end{array} \right.  
\end{align}
where $J_\nu$ is a Bessel function of order $\nu$ and $x=2\alpha^{3/2}\mu/3$. The expression for $|\chi_k|^2$ requires explanation. For $x\ll1$ (i.e., $\alpha\lesssim\mu^{-2/3}$), the asymptotic value is $\chi_k = \mathrm{const.}$, so $|\chi_k|^2\propto\alpha^0$.  For $x\gg1$ (i.e., $\alpha\gtrsim\mu^{-2/3}$), with a choice of phase the asymptotic solution is $\chi_k = A\eta^{-1} \cos(\eta^3\mu/12) = (A/2)\alpha^{-1/2} \cos(\eta^3\mu/12)$, where $A$ is a constant fixed by the evolution of $\chi_k$ to $\IIIMD$. This implies $|\chi_k|^2= (A^2/4)\alpha^{-1}\cos^2(\eta^3\mu/12)$.  Using that solution, there are two terms in $\chi_k'$ (where prime denotes $d/d\eta$): the first term is $- A \eta^{-2}\cos(\eta^3\mu/12)$ and the second term is $- (A/4) \eta\mu\sin(\eta^3\mu/12)$.  The late-time solution in $\IIIMD$ is what we are interested in for GPP, so since the first term in $\chi_k'$ rapidly decays compared to the second term it can be neglected, and $|\chi_k'|^2 = (A^2/4)\alpha\mu^2\sin^2(\eta^3\mu/12)$. The final expression we will use here is that at late time $\omega=\alpha\mu$.  Now $|\chi_k|^2$ and $|\chi_k'|^2$ enter the expression for $n_k$ as:
\begin{align}
2\pi^2 n_k = k^3\left[\frac{1}{\omega_k}\left(\half\omega_k^2 |\chi_k|^2 + \half |\chi_k'|^2\right)-\half\right] \simeq \frac{1}{2}\frac{k^3}{\alpha\mu}\left(\alpha^2\mu^2 |\chi_k|^2 + |\chi_k'|^2\right) = \frac{k^3}{8}\mu A^2 \per 
\end{align}
(For $\mu<1$ and $k<1$ we can ignore the last $\half$ in the first equality.)  With a slight abuse of notation, we will write
\begin{align}\label{eq:nkdef}
4\pi^2n_k = k^3\alpha\mu |\chi_k(\infty)|^2 \com
\end{align}
where it is understood that in this expression one should ignore the oscillatory term in $|\chi_k|^2$.   There is a physical significance to the different regions of $\alpha$ in \eref{eq:chikinIIIMD}, delineated by $\alpha=\mu^{-2/3}$.  From \tref{table:aetaR}, in MD $h=\alpha^{-3/2}$, so $H=m$ corresponds to $\alpha=\mu^{-2/3}$.  If $H>m$, $|\chi_k|^2$ is constant (Hubble drag), while if $H<m$, $|\chi_k|^2$ oscillates with amplitude damping as $\alpha^{-1}$. 

\end{enumerate}

\begin{figure}[hp]
\begin{center}
\includegraphics[width=.95\textwidth]{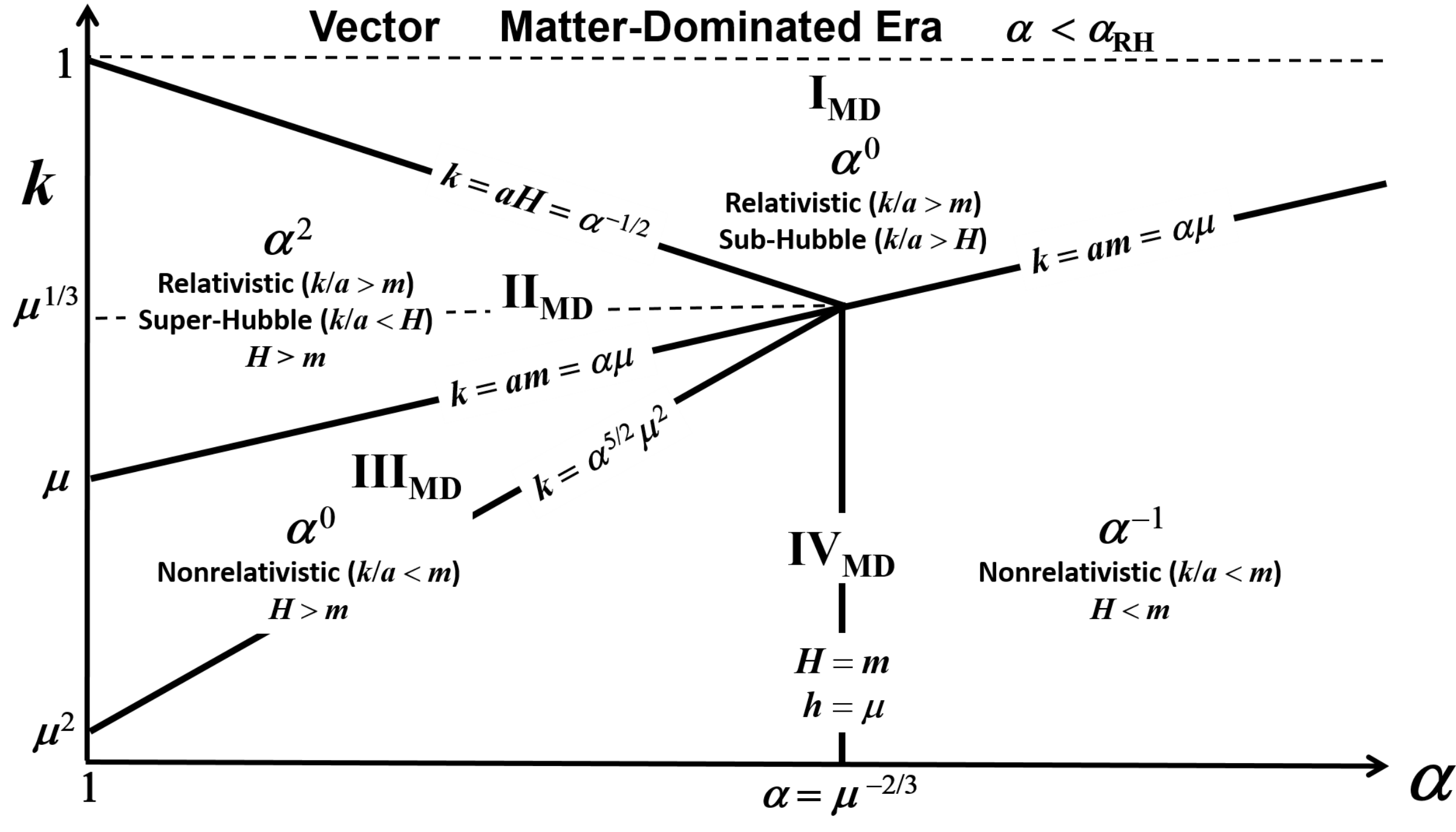}\\
\vspace*{0.25in}
\includegraphics[width=0.95\textwidth]{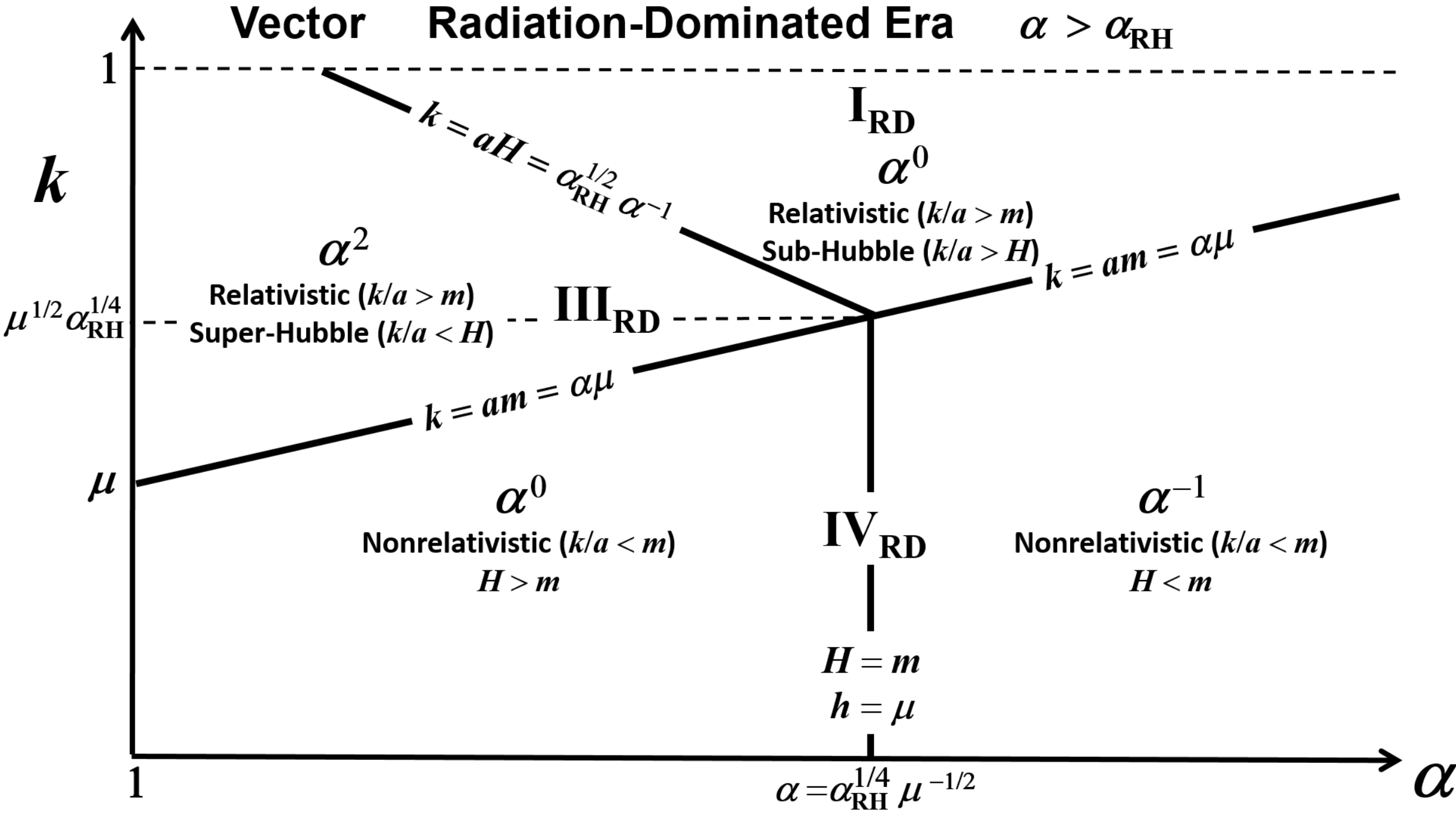}\\
\caption{Regions in the $\alpha$--$k$ plane for evolution in the MD phase (upper) and the RD phase (lower).  In different regions we indicate the scalings of $|\chi_k|^2$with $\alpha$.  Note that $k = \mu^{1/2}\alphaRH^{1/4}$ will be larger than unity (hence off the graph) if $\alphaRH>\mu^{-2}$. In the RD phase, $\alphaRH$ will be smaller (larger) than $\alphaRH^{1/4}\mu^{-1/2}$ if $\alphaRH$ is smaller (larger) than $\mu^{-2/3}$. For both MD and RD, the line denoted $k=\alpha\mu$ is the line delineating the nonrelativistic ($k<am$) and the relativistic ($k>am$) regions. In MD the line $k=\alpha^{-1/2}$ and in RD the line $k=\alpha^{-1}\alphaRH^{1/2}$ is the line denoting $k=aH$; above the lines the mode is sub-Hubble-radius ($k>aH$), while below the line the mode is super-Hubble-radius ($k<aH$). The values of $\alpha=\mu^{-2/3}$ for MD and $\alpha = \alphaRH^{1/4}\mu^{-1/2}$ for RH are the values of $\alpha$ when $H=m$.  To the left of the lines $H>m$, and to the right $H<m$. \label{fig:MD_RD_vector}}
\end{center}
\end{figure}

The results  for MD are also summarized in \tref{table:solutions}.  Various regions and the scaling with $\alpha$ for MD are indicated on the top panel of \fref{fig:MD_RD_vector}.  Also indicated in the figure is the physical significance of various region: relativistic for $k/a>m$, nonrelativistic for $k/a<m$, super-Hubble-radius for $k/a<H$, and sub-Hubble radius for $k/a>H$.  In words: in the relativistic super-Hubble region $|\chi_k|^2\propto \alpha^2$; in the relativistic sub-Hubble region $|\chi_k|^2\propto \alpha^0$; in the nonrelativistic region $|\chi_k|^2\propto \alpha^0$ if $H>m$ and $|\chi_k|^2\propto \alpha^{-1}$ if $H<m$.

%============
%============

\begin{enumerate}\setcounter{enumi}{8} 

\item $\IRD$:  In this relativistic sub-Hubble region the wave equation is\footnote{For Region $\IRD$, $k>\alphaRH^{1/4}\mu^{1/2}$ (see \fref{fig:MD_RD_vector}), so $k^2\gtrsim\alphaRH\mu^2/k^2$.} 
\begin{align}
\chi_k'' + \left(k^2+\dfrac{3\alphaRH\mu^2}{k^2}\right)\chi_k \simeq \chi_k'' + k^2\chi_k = 0 & \Longrightarrow \chi_k = c_1\cos(k\eta)+c_2\sin(k\eta) \nn
& \Longrightarrow |\chi_k|^2 \propto \left\{ 
\begin{array}{ll} \alpha^2 & \alpha<\alphaRH^{1/2}k^{-1} \\ \alpha^0 & \alpha>\alphaRH^{1/2}k^{-1} \per \end{array} 
\right.
\end{align}
Note from \tref{table:aetaR} that in RD $k\eta \simeq k\alphaRH^{-1/2}\alpha$. Thus, for $\alpha<\alphaRH^{1/2}k^{-1}$ the argument of the trigonometric functions is much less than unity and upon expansion yields for the growing mode $\chi_k\simeq c_1\alpha$; hence, $|\chi_k|^2\propto \alpha^2$.  If $\alpha>\alphaRH^{1/2}k^{-1}$ the solution will be an oscillation in $\alpha$ with frequency $k\alphaRH^{-1/2}$ and constant amplitude.

\item $\IIIRD$: In this relativistic super-Hubble region we find
\begin{align}
\chi_k'' + \alphaRH\mu^2\eta^2\chi_k \simeq \chi_k'' + k^2\chi_k = 0 & \Longrightarrow \chi_k=c_1D_{-1/2}[(1+i)x] + c_2D_{-1/2}[(-1+i)x] \nn
& \Longrightarrow |\chi_k|^2 \propto \left\{ 
\begin{array}{ll}\alpha^0 & \alpha \lesssim \alphaRH^{1/4}\mu^{-1/2} \\[1ex] \alpha^{-1} & \alpha \gtrsim \alphaRH^{1/4}\mu^{-1/2} \com \end{array}
\right.
\end{align}
where $D_{-1/2}$ is a parabolic cylinder function and $x = \alpha\alphaRH^{-1/4}\mu^{1/2}$. Expansion of the parabolic cylinder functions for large and small $x$ leads to the indicated scaling of $|\chi_k|^2$ with $\alpha$.

\item $\IVRD$:  In the final nonrelativistic region the equation of motion and solutions are given by
\begin{align}
\chi_k''+ \dfrac{3k^2}{\mu^2\alphaRH^2}\dfrac{1}{\eta^4}\chi_k=0 & \Longrightarrow \chi_k=c_1\ \eta\ \cos \left(\dfrac{\sqrt{3}k}{\alphaRH\mu\eta}\right) + c_2\ \eta\  \sin\left(-\dfrac{\sqrt{3}k}{\mu\alphaRH\eta}\right)\nn
& \Longrightarrow |\chi_k|^2 \propto \alpha^2 \per
\end{align}
By way of explanation, the argument of the trigonometric functions is approximately $k/\alphaRH^{1/2}\mu\alpha$. In $\IVRD$, $k<\alpha\mu$ and $\alphaRH>1$, so the argument of the trigonometric functions are small, and $|\chi_k| \propto \eta \propto \alpha^1$.

\end{enumerate}

The results for RD are also summarized in \tref{table:solutions}.  Various regions and the scaling with $\alpha$ for RD are indicated on the bottom panel of \fref{fig:MD_RD_vector}.  Also indicated in the figure is the physical signifiance of various region: relativistic for $k/a>m$, nonrelativistic for $k/a<m$, super-Hubble-radius for $k/a<H$, and sub-Hubble radius for $k/a>H$.  In words: in the relativistic super-Hubble region $|\chi_k|^2\propto \alpha^2$; in the relativistic sub-Hubble region $|\chi_k|^2\propto \alpha^0$; in the nonrelativistic region $|\chi_k|^2\propto \alpha^0$ if $H>m$ and $|\chi_k|^2\propto \alpha^{-1}$ if $H<m$.

The evolution of the modes in dS, MD, and RD are the same in the various physical regions.  There are however some differences between MD and RD.  Firstly, the demarcations between relativistic super-Hubble and relativistic sub-Hubble are different values of $k$.  Secondly, the values of $\alpha$ for $H=m$ differ.  Finally, the values of $k$ where $aH=am$ differ.

An example of the evolution of $|\chi_k|^2$ with $a$ is shown in \fref{fig:compscalarvector}, where it is compared to the evolution of a minimal scalar for the same values of $k$ and $\mu$.  The important difference between minimal-scalar and vector evolution is in the region $k/m < a \lesssim H_e^{1/3}m^{2/3}$.  In this region $|\chi|^2$ grows as $a^2$ for a minimal scalar and is constant for a vector.  Thus, the final result will be a factor of $\left[(k/m)/(H_e^{1/3}m^{2/3})\right]^2=k^2/\mu^{2/3}$ smaller.  For $k=10^{-3}$ and $\mu= 10^{-2}$ illustrated in \fref{fig:compscalarvector}, $k^2/\mu^{2/3}=2.15\times 10^{-5}$, which agrees well with the final ratio of $|\chi_k|^2$.

%==================
\subsection{Evolution of the modes} \label{section:evolution}
%==================

Now we will start with a $k$-mode deep in the de Sitter era in the Bunch--Davies vacuum and follow $|\chi_k|^2$ until it reaches the nonrelativistic region with $m>H$.  In this region $|\chi_k|^2$ oscillates with amplitude decreasing as $\alpha^{-1}$.  In the region the evolution is adiabatic and one can sensibly defining a number density of particles resulting from GPP.  This will be the asymptotic behavior of $|\chi_k|^2$, and thereafter $\alpha|\chi_k|^2$ will remain constant.

%===================
\subsubsection{de Sitter evolution} \label{section:evolution_deSitter}
%===================

We first consider the evolution of $|\chi_k|^2$ in the de Sitter era.  As $\alpha\to0$, we will assume Bunch-Davies vacuum and take as initial conditions
\begin{align}
|\chi_k(\alpha=0)|^2=\frac{1}{2k} \per
\end{align}
Starting with those initial conditions we can follow the evolution of $|\chi_k|^2$ easily by referring to \fref{fig:dS_vector}.

Consider two cases for the evolution of $|\chi_k|^2$ in the dS era:
\begin{enumerate}
\item $1>k>\mu$.  The mode begins in the Bunch-Davies vacuum and remains constant until $\alpha=k/\sqrt{2}$.  Then it grows as $\alpha^2$ in $\IIdS$ until the end of inflation.  So at $\alpha=1$,
\begin{align}
|\chi_k(\alpha=1)|^2 = \frac{1}{2k} \left(\frac{1}{k/\sqrt{2}}\right)^2 =\frac{1}{k^3} \qquad\qquad (1>k>\mu) \per
\end{align}
\item $\mu >k > 0$: Again, the mode begins in the Bunch-Davies vacuum and remains constant until $\alpha=k/\sqrt{2}$.  Then it grows as $\alpha^2$ until it crosses $\alpha=k/\mu$, after which it remains constant until the end of inflation.  At $\alpha=1$,
\begin{align}
|\chi_k(1)|^2 = \frac{1}{2k} \left(\frac{k/\mu}{k/\sqrt{2}}\right)^2 = \frac{1}{k\mu^2} \qquad\qquad (\mu>k>0) \per
\end{align}
\end{enumerate}

In conclusion, the mode amplitudes (squared) at the end of inflation are
\begin{align}\label{eq:enddeSitter}
|\chi_k(1)|^2 = \left\{ \begin{array}{ll} \dfrac{1}{k^3} & \qquad\quad (1>k>\mu) \\[2ex]
\dfrac{1}{k\mu^2} & \qquad\quad (\mu>k>0) \per \end{array}  \right.
\end{align}
This result agrees with calculations by other authors such as Refs.~\cite{Ema:2019yrd,Ahmed:2020fhc}.

One caveat is that we have assumed $R$ is constant in dS.  In a slow-roll model typically $R$ grows as a logarithm in $\alpha$ as $\alpha\to0$.  We will discuss a correction for this later.

%===================
\subsubsection{Matter-dominated evolution} \label{section:evolution_MD}
%===================

Now consider the evolution of $|\chi_k|^2$ in the matter-dominated era until the evolution to the nonrelativistic, sub-Hubble-radius region. This amounts to following the evolution past $\alpha=\mu^{-2/3}$ (see \fref{fig:MD_RD_vector}) assuming the mode reaches the nonrelativistic, $H<m$ region before reheating. We will describe this possibility as the late-reheating case. We will denote this asymptotic value of $|\chi_k|^2$ as $|\chi_k(\alpha\to\infty)|^2$.

There will be three cases, depending on the value of $k$. Again, with the help of the upper panel in \fref{fig:MD_RD_vector} we can follow the evolution through the MD era.  Our goal is to find the value of $|\chi_k(\infty)|^2$, which will be used to calculate $n_k$.

\begin{enumerate}

\item $1>k>\mu^{1/3}$: At $\alpha=1$ the mode enters MD in the relativistic-super-Hubble region and grows as $\alpha^2$ until it crosses into the relativistic-sub-Hubble region at $\alpha=k^{-2}$ and remains constant until it becomes nonrelativistic at $\alpha=k/\mu$, after which it damps as $\alpha^{-1}$. Putting things together,
\begin{align}
|\chi_k(\infty)|^2 = \frac{1}{k^3}\ \left(\frac{k^{-2}}{1}\right)^2\ \frac{k/\mu}{\alpha} = \frac{1}{k^3\mu\alpha}\ \frac{1}{k^3} \per
\end{align}
where the first factor of $k^{-3}$ is the value of $|\chi_k(\alpha=1)|^2$ for $k>\mu$ from \eref{eq:enddeSitter}.  Using \eref{eq:chikinIIIMD}, 
\begin{align}
4\pi^2 n_k = k^3\mu\alpha \ |\chi_k(\infty)|^2  = \frac{1}{k^3} \qquad (1>k>\mu^{1/3}) \per
\end{align}
The expression relating $n_k$ and $|\chi_k(\infty)|^2$ will be used often. 

\item $\mu^{1/3} > k > \mu$: In this range of $k$ the mode again enters MD in the relativistic-super-Hubble region and evolves as $\alpha^2$.  Then, when $\alpha=k/\mu$ it enters the nonrelativistic, $H>m$ region, after which it remains constant until it crosses $\alpha=\mu^{2/3}$, then it damps as $\alpha^{-1}$.   Gluing together the pieces of evolution,
\begin{align}
4\pi^2 n_k & = k^3\mu\alpha\ \frac{1}{k^3}\ \left(\frac{k/\mu}{1}\right)^2\ \frac{\mu^{-2/3}}{\alpha}  \nn
4\pi^2 n_k & = \frac{k^2}{\mu^{5/3}} \qquad (\mu^{1/3} > k > \mu) \per
\end{align}

\item $\mu > k > 0$:  For this final case the mode enters MD through the nonrelativistic $H>m$ region and remains constant until it enters crosses into the $H<m$ region at $\alpha=\mu^{-2/3}$ when it begins damped oscillations. Thus,
\begin{align}
4\pi^2n_k & = k^3\mu\alpha \ \frac{1}{k\mu^2} \frac{\mu^{-2/3}}{\alpha}  \nn
4\pi^2n_k & = \frac{k^2}{\mu^{5/3}} \qquad (\mu > k > 0) \per
\end{align}
Note than now we have used $|\chi_k(\alpha=1)|^2=(k\mu^2)^{-1}$ for $\mu>k$ as in \eref{eq:enddeSitter}.  The evolution of $|\chi_k|^2$ for a value of $k$ in this range was illustrated in \fref{fig:compscalarvector}.

\end{enumerate}

\begin{figure}[p]
\begin{center}
\includegraphics[width=0.95\textwidth]{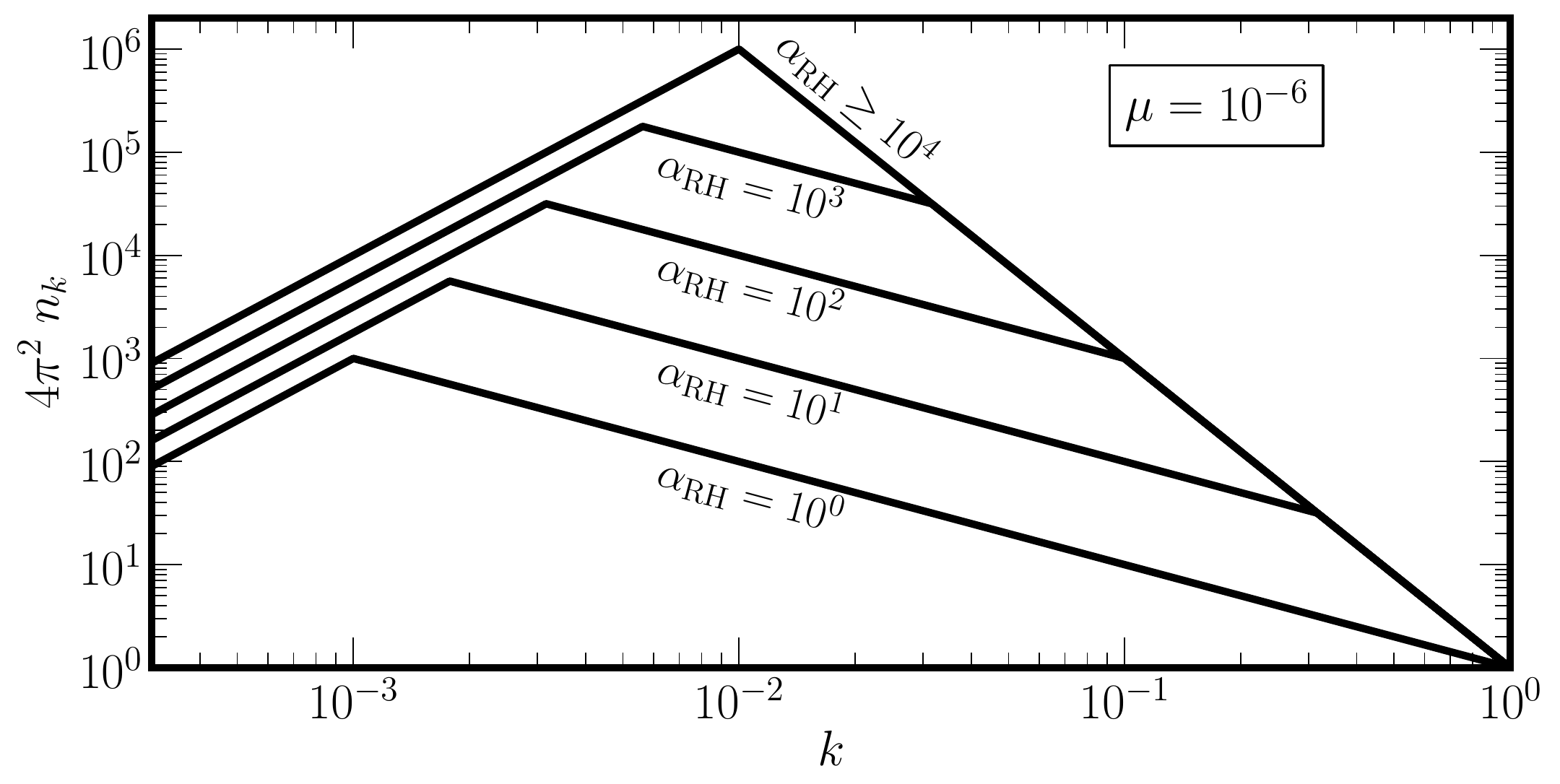}\\
\includegraphics[width=0.95\textwidth]{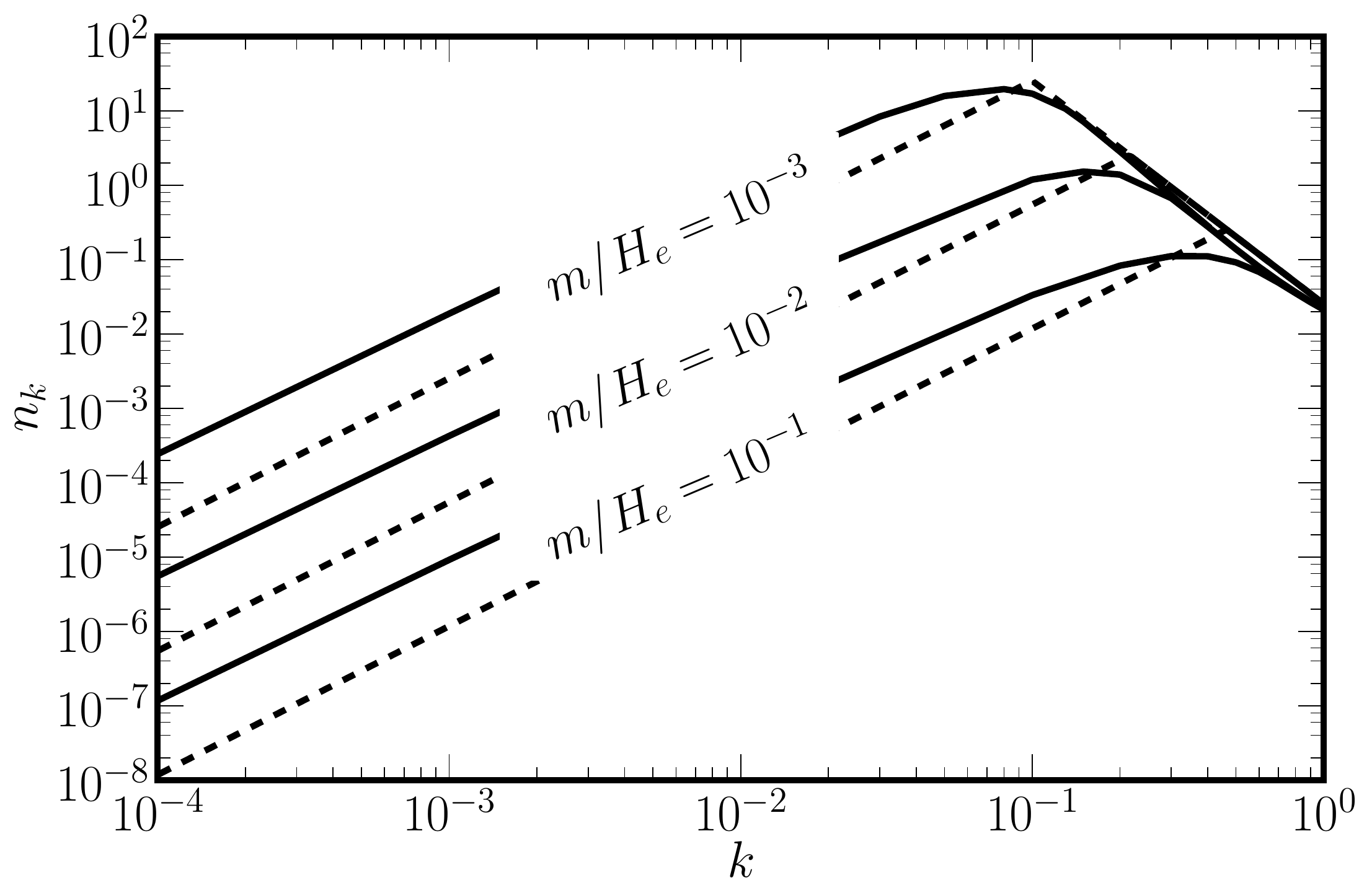}\\
\caption{Top panel: The final spectrum as a function of $k$ for late reheating ($\alphaRH\mu^{2/3}>1$; for $\mu=10^{-6}$ this implies $\alphaRH \geq 10^4$) using  \eref{eq:vectordensityMD}, for early reheating ($1>\alphaRH\mu^{2/3}>\mu^{2/3}$) using \eref{eq:onetotal}, and for immediate reheating ($\alphaRH=1$) using \eref{eq:instant}.   Bottom panel: the final value of $n_k$ for $m/H_e=10^{-3}$, $10^{-2}$, and $10^{-1}$ assuming $\alphaRH\mu^{2/3} > 1$.  The solid curves are numerical results for the chaotic model while the dashed curves are the analytic approximation of \eref{eq:nk_analytic_vector}.  At low $k$ they differ by a factor of $\sim 10$, which is explained in the text. \label{fig:totalspectrum}}
\end{center}
\end{figure}

The conclusion is that for $\alphaRH>\mu^{-2/3}$,
\begin{align}\label{eq:nk_analytic_vector}
n_k = \dfrac{1}{4\pi^2}\left\{ \begin{array}{ll}
k^{-3}      	& \qquad (1>k>\mu^{1/3}) \\[2ex]
k^2\mu^{-5/3}	& \qquad (\mu^{1/3} > k > 0) \per
\end{array} \right.
\end{align}
Of course the values are equal at $k=\mu^{1/3}$.  This scaling is shown in \fref{fig:totalspectrum} in cartoon form as the dashed curve in the top panel for a particular choice of $\mu$, and compared to the numerical results for three values of $\mu$ in the lower panel.  Note that the spectrum is rather peaked around $k=\mu^{-1/3}$.  We can find the total number density by integrating \eref{eq:nk_analytic_vector}:
\begin{align}\label{eq:vectordensityMD}
na^3 = \int_0^1\frac{dk}{k} \ n_k = \dfrac{1}{4\pi^2} \ \dfrac{5}{6}\left(\dfrac{1}{\mu}-\frac{2}{5}\right)  \per
\end{align}
Modes of higher $k$ are damped and won't contribute significantly to $na^3$. Also, recall that we are only considering $\mu<1$ since higher-mass modes are also damped.

%============
%============

In the infrared the $k$-dependence of the analytic results is $k^{2}$, while the numerical results for the chaotic model are better fit by a dependence of $k^{1.8}$.  This discrepancy is due to the fact that the scalar curvature $R$ is not constant in the chaotic model, but increases as one goes further back in inflation.  The numerical result $k^{1.8}$ is for the chaotic model, and a different inflation model may give a different scaling.   The value of the integrated spectrum will not depend much on the exact  infrared behavior so long as $n_k\to0$ as $k\to0$. This is problematic for a minimally-coupled scalar, but no problem for the vector, which has a blue spectrum.  The infrared dependence  will have a larger effect on the isocurvature component as well as nongaussianities.  So long as the spectrum decreases in the infrared faster than $k$, isocurvature issues should not arise \cite{Graham:2015rva}.

Since the scaling of $n_k$ with $k$ in the infrared leads to a convergent result for $na^3 = \int n_k d\ln k$, the infrared behavior does not much affect the total number density; rather, the total number density depends on  the value of $n_k$ around the peak at $k\simeq\mu^{1/3}$.  From \eref{eq:nk_analytic_vector}, the peak value scales as $n_k(k=\mu^{1/3})\propto \mu^{-1}=H_e/m$.  This implies that the contribution to the mass density, proportional to $m\, n\, a^3$, is roughly independent of $m$!  This will be discussed in the next section.

%===================
\subsubsection{Radiation-dominated evolution} \label{section:evolution_RD}
%===================

Now consider the effects of reheating, which is important if reheating occurs before the mode reaches the nonrelativistic, sub-Hubble-radius region. We will call this the early-reheating case.  In the early reheating case we must consider RD evolution.

The evolution of $|\chi_k|^2$ in the RD era is shown in \fref{fig:MD_RD_vector}.  It is useful to refer to the figure when discussing the evolution through reheating.  First, we establish a hierarchy of inequalities for $\alphaRH<\mu^{-2/3}$:
\begin{align}
1 > \alphaRH^{-1/2} > \mu^{1/3} > \alphaRH^{1/4}\mu^{1/2} > \alphaRH\mu > \alphaRH^{5/2}\mu^2 > \mu^2\alphaRH^{-1/2} > 0 \per
\end{align} We will again study the evolution for various ranges of $k$.

\begin{enumerate}

\item $1 > k > \alphaRH^{-1/2}$:  In this range the mode enters the MD region as relativistic, super-Hubble and evolves as $\alpha^2$ until it crosses into the relativistic sub-Hubble region at $\alpha=k^{-2}$.  Then it evolves as a constant,  reheating occurs in this region and the mode continues to evolve as a constant  until $\alpha=k/\mu$ when it enters the nonrelativistic $H<m$ region and thereafter damps as $\alpha^{-1}$.  The final value of $n_k$ will be
\begin{align}\label{eq:oneone}
4\pi^2n_k & = k^3\mu\alpha \ \dfrac{1}{k^3} \ \left(\dfrac{k^{-2}}{1}\right)^2\ \dfrac{k/\mu}{\alpha} \nn
4\pi^2n_k & = k^{-3} \qquad (1 > k > \alphaRH^{-1/2})\per
\end{align}
The factor of $k^3\mu\alpha$ converts $|\chi_k(\infty)|^2$ to $n_k$, see \eref{eq:nkdef}, and the factor of $k^{-3}$ is $|\chi_k(1)|^2$ from \eref{eq:enddeSitter}.

\item $\alphaRH^{-1/2} > k > \mu^{1/3}$: In this region the evolution in MD begins as above, but reheating at $\alpha=\alphaRH$ occurs in the relativistic super-Hubble region before the mode crosses $\alpha=k^{-2}$.  The mode then continues to grow as $\alpha^2$ in the relativistic super-Hubble region of RD until $\alpha=\alphaRH^{1/2}/k$.  Then it evolves through the relativistic sub-Hubble region as a constant until $\alpha=k/\mu$ and damped oscillations commence.  This results in
\begin{align}\label{eq:onetwo}
4\pi^2n_k & = k^3\mu\alpha \ \dfrac{1}{k^3} \ \left(\dfrac{\alphaRH^{1/2}/k}{1}\right)^2\ \dfrac{k/\mu}{\alpha} \nn
4\pi^2n_k & = k^{-1} \alphaRH \qquad (\alphaRH^{-1/2} > k > \mu^{-1/3} )\per
\end{align}

\item $\mu^{1/3} > k > \alphaRH^{1/4}\mu^{1/2}$: Since $k>\alphaRH^{1/4}\mu^{1/2}$, $k$ will satisfy $k>\alphaRH\mu$.  This implies reheating will occur while the mode is in the MD relativistic super-Hubble region before it crosses $k=\alpha\mu$.  After reheating the mode will continue to grow as $\alpha^2$ in the RD relativistic super-Hubble region until $\alpha=\alphaRH^{1/2}/k$.  It then remains constant until $\alpha=k/\mu$ and begins damped oscillations.  Thus, the final result will be 
\begin{align}\label{eq:onethree}
4\pi^2n_k & = k^3\mu\alpha \ \dfrac{1}{k^3} \ \left(\dfrac{\alphaRH^{1/2}/k}{1}\right)^2 \ \dfrac{k/\mu}{\alpha} \nn 
4\pi^2n_k & = k^{-1}\alphaRH\qquad (\mu^{-1/3} > k > \alphaRH^{1/4}\mu^{1/2})\per
\end{align}

\item $\alphaRH^{1/4}\mu^{1/2} > k > \alphaRH\mu$:  The mode enters MD in the relativistic super-Hubble region scaling as $\alpha^2$ as previously.  It reheats before becoming nonrelativistic and continues to evolve in RD as $\alpha^2$ until $\alpha=k/\mu$ when it enters the nonrelativistic $H>m$ region as remains constant until it crosses $\alpha=\alphaRH\mu^{-1/2}$ and starts damped oscillations.  This leads to the result
\begin{align}\label{eq:onefour}
4\pi^2n_k & = k^3\mu\alpha \ \dfrac{1}{k^3} \ \left(\dfrac{k/\mu}{1}\right)^2 \  \dfrac{\alphaRH^{1/4}\mu^{-1/2}}{\alpha} \nn
4\pi^2n_k & = k^2\alphaRH^{1/4}\mu^{-3/2} \qquad (\alphaRH^{1/4}\mu^{1/2} > k > \alphaRH\mu) \per
\end{align}  

\item $\alphaRH\mu > k > \mu$: Now the mode scales as $\alpha^2$ until it becomes nonrelativistic in MD at $\alpha=k/\mu$.  Then it is constant in the nonrelativistic $H>m$ region before and after reheating until it becomes nonrelativistic and commences damped oscillation.   Therefore,
\begin{align}
4\pi^2n_k & = k^3\mu\alpha \ \dfrac{1}{k^3} \  \left(\frac{k/\mu}{1}\right)^2 =\frac{1}{k\mu^2} \ \dfrac{\alphaRH^{1/4}\mu^{-1/2}}{\alpha} \nn
4\pi^2n_k & =  k^2\alphaRH^{1/4}\mu^{-3/2} \qquad(\alphaRH\mu>k>\mu) \per
\end{align}

\item $\mu > k > 0$: The mode now enters MD in the nonrelativistic $H>m$ region where the more remains constant, and will remain so after reheating until $\alpha=\alphaRH^{1/4}\mu^{-1/2}$ and damped oscillations begin.  This leads to
\begin{align}\label{eq:onesix}
4\pi^2n_k & = k^3\mu\alpha \ \dfrac{1}{k\mu^2} \ \dfrac{\alphaRH^{1/4}\mu^{-1/2}}{\alpha} \nn
4\pi^2n_k & = k^2\alphaRH^{1/4}\mu^{-3/2} \qquad (\mu> k > 0) \per
\end{align}

\end{enumerate}
 
Assembling the results from \eref{eq:oneone} through \eref{eq:onesix} leads to the final result 
\begin{align}\label{eq:onetotal}
n_k = \dfrac{1}{4\pi^2} \left\{ \begin{array}{ll}
k^{-3} & \qquad 1>k>\alphaRH^{-1/2} \\[2ex]
k^{-1}\alphaRH & \qquad \alphaRH^{-1/2} > k > \alphaRH^{1/4}\mu^{1/2} \\[2ex]
k^2\alphaRH^{1/4}\mu^{-3/2} & \qquad \alphaRH^{1/4}\mu^{1/2} > k > 0 \per
\end{array}\right.
\end{align}
This result is shown in graphical form by the dashed curve in the top panel of \fref{fig:totalspectrum} for the choice $\mu=10^{-6}$ and $\alphaRH=10^3$, which satisfies the condition $\alphaRH<\mu^{-2/3}$.

Comparing the two spectra in the upper panel of \fref{fig:totalspectrum}, we see that the spectrum peaks at a smaller value of $k$ if $\alphaRH>\mu^{-2/3}$.  We also see that the maximum value of $n_k$ is smaller if $\alphaRH<\mu^{-2/3}$.

Since the spectrum is convergent in the IR, we can again integrate the spectrum of \eref{eq:onetotal} to yield $na^3$:
\begin{align}\label{eq:vectordensityRD}
na^3 = \dfrac{1}{4\pi^2}\int_0^1{\frac{dk}{k}\ n_k} = \dfrac{1}{4\pi^2}\ \left[  \dfrac{3}{2}\alphaRH^{3/4}\mu^{-1/2}\left( 1 - \dfrac{4}{9}\alphaRH^{3/4}\mu^{1/2} \right) - \dfrac{1}{3}\right] \per
\end{align}
Since we are assuming $\alphaRH\mu^{2/3}<1$, the second term in the parenthesis is less than unity and $\alphaRH^{3/4}\mu^{-1/2}>1$.   Note that if $\alphaRH\mu^{2/3}=1$, we recover the result of \eref{eq:vectordensityMD}. Important for the next section is that to leading order in $\alphaRH\mu^{2/3}$ the result for $na^3$ is proportional to $\alphaRH^{3/4} \propto \TRH^{-1}$ (see \eref{eq:TRHETC}).  Finally, the ratio of the integrated spectra of early/late reheating is approximately $(\alphaRH\mu^{2/3})^{3/4}<1$.  

In the case of ``immediate'' reheating after inflation ($\alphaRH=1$), \eref{eq:onetotal} becomes
\begin{align}\label{eq:instant}
n_k = \dfrac{1}{4\pi^2} \left\{ \begin{array}{ll}
k^{-1} & \qquad 1 > k > \mu^{1/2} \\[2ex]
k^2\mu^{-3/2} & \qquad \mu^{1/2} > k > 0 \com
\end{array}\right.
\end{align}
and the integrated spectral density yields 
\begin{align}\label{eq:nkinstant}
%n_k = \dfrac{1}{4\pi^2} \left(\frac{H_e}{m}\right)^{1/2} \per
na^3 = \frac{1}{4\pi^2} \left(\frac{3}{2} \sqrt{\frac{H_e}{m}} -1\right) \per
\end{align}
In the special case $\alphaRH = 1$ the result agrees with Graham, Mardon, and Rajendran \cite{Graham:2015rva}.

The lower panel of \fref{fig:totalspectrum} shows the spectrum $n_k$ for several values of the scalar's mass $m$ in units of $H_e$.  
The dashed curves correspond to the analytic approximations discussed above, while the solid curves correspond to a direct numerical solution of the mode equations.  
Notice that the analytic approximations have underestimated the spectrum by a factor of $\sim 10$ at low $k$.  
This can be understood from the evolution of the Hubble parameter during inflation.  
For the numerical work, we assume a chaotic model of inflation with a quadratic inflaton potential.  
In this model the Hubble parameter decreases by a factor of $\sim 10$ between the time of CMB mode generation and the end of inflation.  
Since the modes with smaller $k$ leave the horizon earlier, they probe the larger $H > H_e$, which leads to a larger $n_k$ relative to the analytic approximations that assume $H = H_e$ throughout inflation.  
Nevertheless, we see also from \fref{fig:totalspectrum} that the analytic approximation works well for the modes where the spectrum is peaked, which means that the total abundance $na^3$ can be calculated reliably from the analytic approximations while only introducing an $O(1)$ error.  
For other models of inflation in which the inflation potential is shallower and the $H_\mathrm{inf} / H_e$ is not much larger than $1$, such as the $\alpha$-attractor class of models~\cite{Kallosh:2013yoa} including Starobinsky's $R^2$ inflation~\cite{Starobinsky:1980te}, we expect that our analytic treatment will provide an even better approximation of the spectrum.  

%===================
\begin{table}[th]
\begin{center}
\caption{\label{table:alpharh_regions} Results for $4\pi^2 n_k$ for immediate, early, and late reheating. \label{table:connection}} 
\begin{tabular}{|l|c|c|}\hline
Immediate Reheating & Early Reheating & Late Reheating\\ 
$\alphaRH=1$    & $\mu^{-2/3}>\alphaRH>1$ & $\alphaRH>\mu^{-2/3}$ \\[1ex]\hline & & \\
$ \begin{array}{ll} k^{-1} & (1 > k > \mu^{1/2}) \\[1ex] 
k^2\mu^{-3/2} & (\mu^{1/2} > k > 0) \end{array}$ & 
$\begin{array}{ll}
k^{-3} & (1>k>\alphaRH^{-1/2}) \\[1ex]
k^{-1}\alphaRH & (\alphaRH^{-1/2} > k > \alphaRH^{1/4}\mu^{1/2}) \\[1ex]
k^2\alphaRH^{1/4}\mu^{-3/2} & (\alphaRH^{1/4}\mu^{1/2}) > k > 0 \end{array} $
& $\begin{array}{ll} k^{-3} & (1>k>\mu^{1/3}) \\[1ex]
k^2\mu^{-5/3}	& (\mu^{1/3} > k > 0)\end{array}$ \\ & & \\ \hline
\end{tabular}
\end{center}
\end{table}
%===================

%------------------------------ 
\section{Contribution to the present mass-energy density \label{sec:relic}}
%------------------------------ 

We will be interested in the present number density of particles from GPP.  At late times\footnote{Again, by ``late times'' we mean $|\chi_k^2|$ has evolved to the nonrelativistic $H<m$ region.} the comoving number density $na^3$ is constant, as is the comoving entropy density $sa^3$ after reheating, where $s=(2\pi^2/45)g_*T^3$ is the entropy density.  Here, $g_*$ counts the number of degrees of freedom. We assume that after reheating the expansion rate (squared) is $H^2 = H_e^2\alphaRH/\alpha^4 = \kappa\rho_R/3$, where $\rho_R = (\pi^2g_*/30)T^4$ is the radiation density.   Equating these two expression for $H^2$ in the radiation era and using \eref{eq:TRHETC} to express $\alphaRH$ in terms of $\TRH$ leads to $sa^3=4\Mpl^2/H_e\TRH$. Taking advantage of the fact that $n/s\propto\mathrm{const.}$, the ratio of the present number density of the GPP and the entropy density is
\begin{align}
n_0 =\frac{[na^3]}{[sa^3]}\ s_0 \com
\end{align}
where $s_0\simeq3000\cm^{-3}$ is the present entropy density.  The present mass density is $m n_0$, and expressing it in terms of $\Omega=\rho_0/3H_0^2\Mpl^2$, the result is
\begin{align}\label{eq:omegah2} 
\frac{\Omega h^2}{0.12} = \frac{m}{H_e}\left(\frac{H_e}{10^{12}\GeV}\right)^2\left(\frac{\TRH}{10^9\GeV}\right) \ \frac{\left[na^3\right]}{10^{-5}} \per 
\end{align}

%============
\begin{figure}[htp]
\begin{center}
\includegraphics[width=0.95\textwidth]{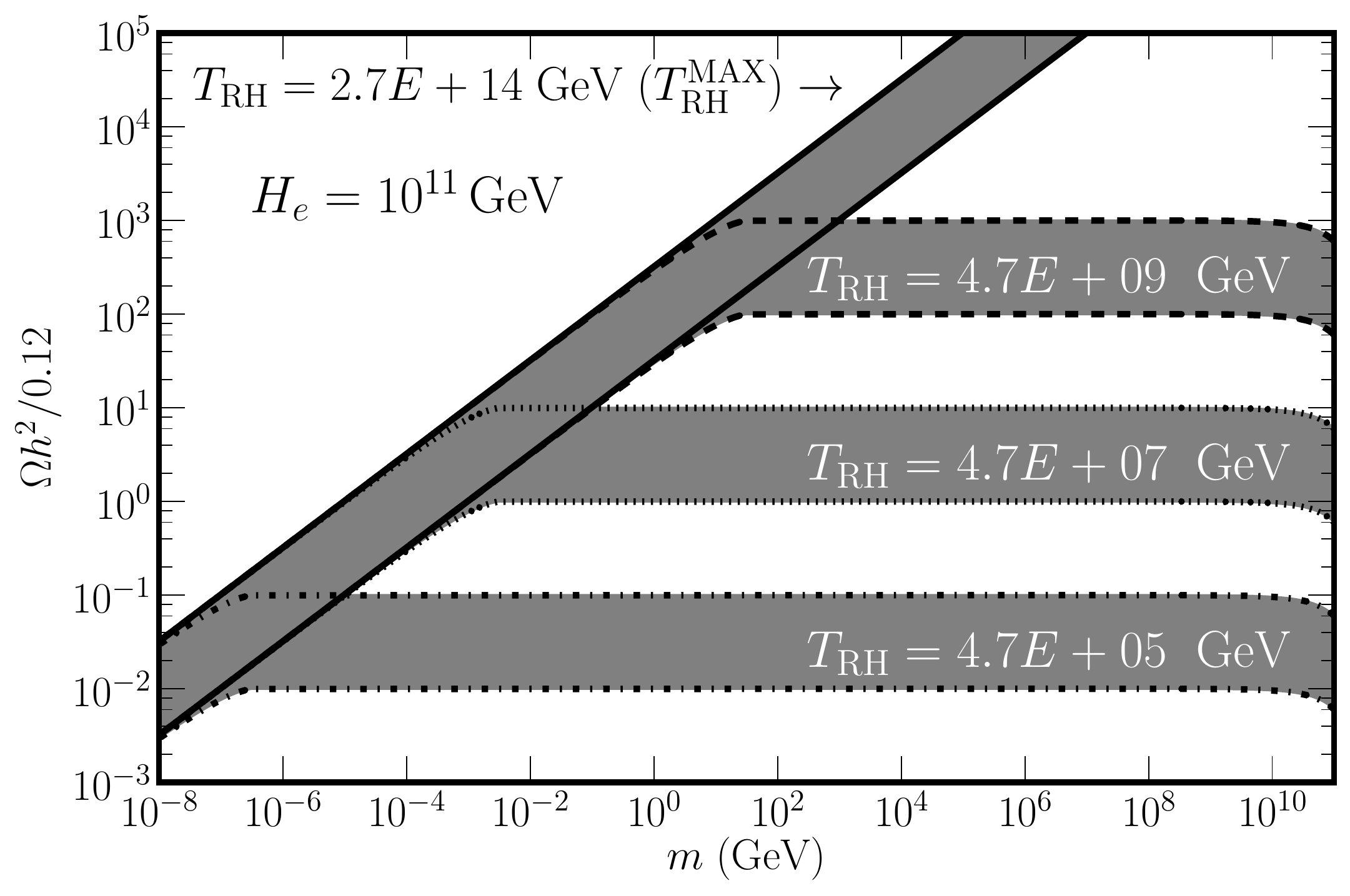}\\
\includegraphics[width=0.95\textwidth]{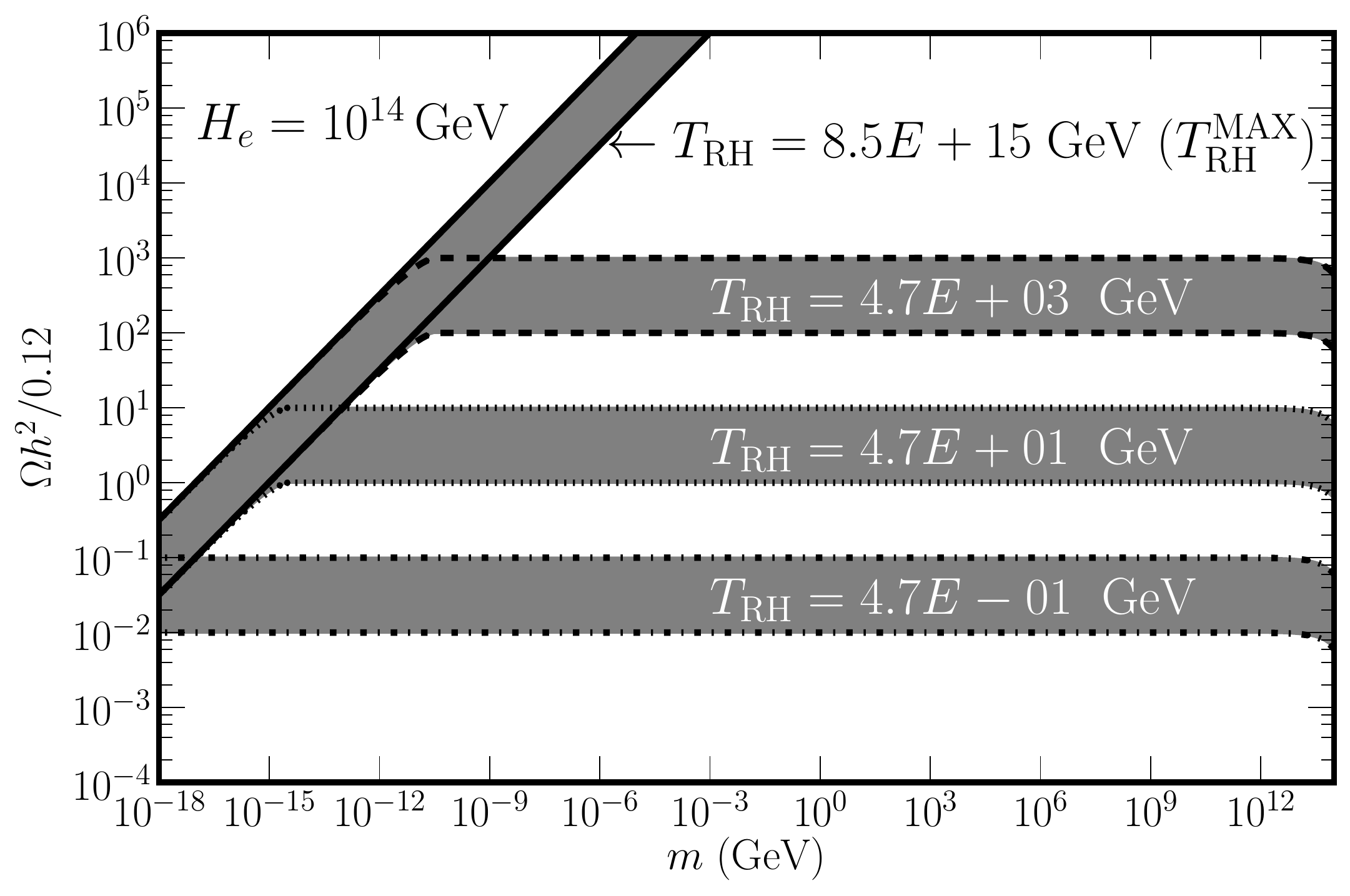}\\
\caption{The relic abundance of longitudinally-polarized dark photon dark matter, $\Omega h^2$, as a function of its mass, $m$, and the reheating temperature, $\TRH$, for two values of the inflationary Hubble scale, $H_e$.  The asymptotic behavior is captured by \erefs{eq:latereheat}{eq:earlyreheat}.  The band illustrates a weak dependence on the inflationary model (for a given $H_e$). 
\label{fig:omegavsm}}
\end{center}
\end{figure}
%============

We now determine the relic abundance $\Omega h^2$ using \eref{eq:omegah2}.  
In the late reheating case,\footnote{As discussed in the previous section, late reheating means that the mode has reached the nonrelativistic $H<m$ region before reheating, and early reheating refers to the case when it reaches the nonrelativistic $H<m$ region after reheating.} our analytic calculation of the comoving number density $na^3$ appears in \eref{eq:vectordensityMD}.  
As we have discussed previously, our numerical calculation that appears in \fref{fig:totalspectrum} indicates that the analytic calculation underestimates the spectrum by a factor of $\sim 10$ at low $k$, and we associated this factor with the assumed chaotic inflation model.  
After accounting for this additional factor, the relic abundance is found to be 
\begin{align}\label{eq:latereheat}
\frac{\Omega h^2}{0.12} = (1\textendash10) \times \left(\frac{H_e}{10^{11}\GeV}\right)^2\left(\frac{\TRH}{5\times10^7\GeV}\right)\left(1-\frac{2}{5}\frac{m}{H_e}\right) \qquad \left(\TRH < 8.4\times10^8\left(\frac{m}{\GeV}\right)^{1/2} \GeV  \right) \com
\end{align}
where the variable prefactor $(1\textendash10)$ accounts for the weak dependence on the inflationary model.  
We have also used \eref{eq:MDRD} to express $\alphaRH$ in terms of $\TRH$.   Of note is the result that (to leading order in $m/H_e$) for late reheating $\Omega h^2$ is independent of $m$. Also, if $H_e\lesssim 10^8\GeV$ and $m<H_e$, the value of $\TRH$ required exceeds the minimum required for late reheating and one cannot have $\Omega h^2=0.12$ for late reheating.  

Now, for early reheating ($\alphaRH\mu^{2/3}<1$) case we use \eref{eq:vectordensityRD} for the value of $na^3$, and to leading order in $\alphaRH\mu^{2/3}$ \cite{Graham:2015ifn,Herring:2019hbe}
\begin{align}\label{eq:earlyreheat}
\frac{\Omega h^2}{0.12} = (1\textendash10) \times  \left(\frac{m}{10^{-6}\eV}\right)^{1/2} \left(\frac{H_e}{10^{14}\GeV}\right)^2  \qquad \left(\TRH > 8.4\times10^8\left(\frac{m}{\GeV}\right)^{1/2} \GeV  \right) \per
\end{align}
Just as the result for late reheating was independent of $m$, to lowest order in $\alphaRH m^{2/3}$ the result for early reheating is independent of $\TRH$.  The same result holds for immediate reheating.  Note that if $H_e\lesssim 10^8\GeV$, then the value of $\TRH$ required exceeds $\TRH^\mathrm{MAX}$.

In general, $\Omega h^2$ depends on three parameters: $m$, $H_e$, and $\TRH$.  In the late-reheating region, $\Omega h^2 \propto H_e^2\TRH$ and is independent of $m$.  In the early-reheating region, $\Omega h^2 \propto H_e^2 m^{1/2}$, and is independent of $\TRH$.    The break in $\Omega h^2$ is at $m=1.4\,(\TRH/10^9\GeV)^2\ \GeV$.  For $m$ greater than this value $\Omega h^2$ is independent of $m$, and for smaller $m$ it is independent of $\TRH$ and decreases as $m^{1/2}$.  The final summary of the results are given in \tref{table:finalsummary}.  Note that the result for immediate reheating agrees with the analysis of GMR \cite{Graham:2015rva}.  

%------------------------------ 
\section{Conclusions \label{sec:conc}}
%------------------------------ 

%============
\begin{figure}[t]
\begin{center}
\includegraphics[width=0.95\textwidth]{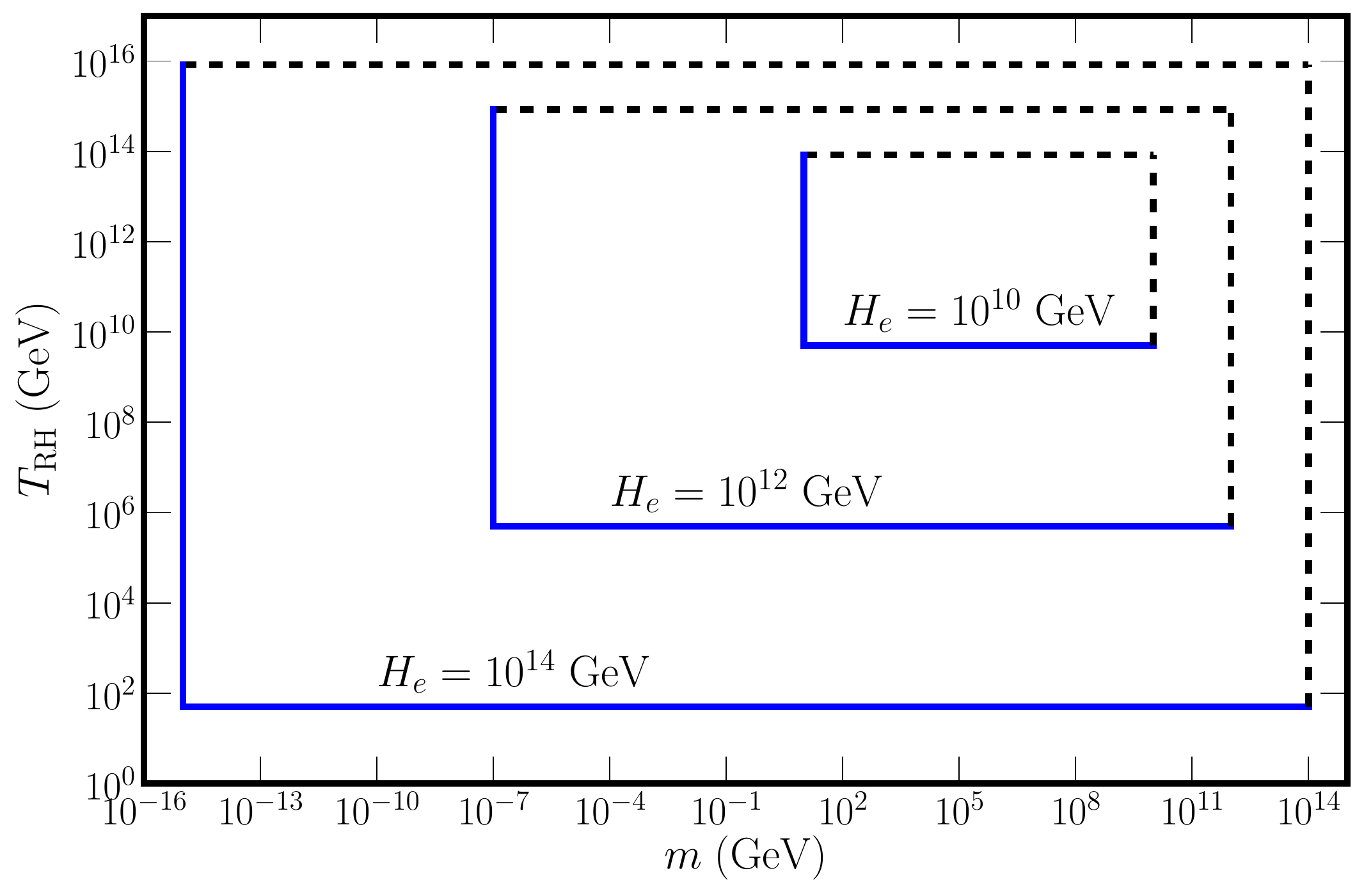}\\
\caption{\label{fig:constraint_plot}
The solid lines are the values of $(m,\TRH)$ that result in $\Omega h^2=0.12$ for the indicated values of $H_e$.  The horizontal solid line is for late reheating while the vertical solid line is for early reheating. As mentioned in the discussion after \eref{eq:latereheat}, if $H_e\lesssim10^8\GeV$ there is no late-time reheating solution that results in $\Omega h^2=0.012$, while as discussed after \eref{eq:earlyreheat} if $H_e\lesssim 10^8$ there is no early-reheating solution that gives $\Omega h^2=0.12$. The vertical dashed lines indicate $m=H_e$; GPP is suppressed for $m>H_e$. The horizontal dashed lines indicate the maximum reheat temperature $\TRH^\mathrm{MAX}$ for a given $H_e$ allowed by energy conservation.  Values of $(m,\TRH)$ inside the rectangles are forbidden since they would result in $\Omega h^2 > 0.12$.  }
\end{center}
\end{figure}
%============

To conclude, let us first summarize the work that was presented here.  Our goal is a calculation of the production of spin-1 dark matter particles during the epoch of inflation and reheating through the phenomenon of gravitational particle production.  

In earlier work by Graham et al.\ \cite{Graham:2015rva}, the spectrum and relic abundance of gravitationally-produced spin-1 dark matter was calculated under the assumption that reheating occurs instantaneously.  This is an effective approximation for ultra-light dark-photon dark matter, but it is not applicable when the dark photon mass becomes larger, $m \gtrsim (1 \GeV) (T_\mathrm{RH} / 10^9 \GeV)^2$.  Here we generalize and extend the analysis by allowing for a finite duration of reheating, which is assumed to be a matter-dominated phase; see also Refs.~\cite{Ema:2019yrd,Ahmed:2020fhc} that present closely related analyses.  We calculate the vector field's mode functions during inflation and reheating both numerically (assuming a quadratic inflaton potential $V \propto \phi^2$) and analytically, finding excellent agreement between these two approaches.  For the analytic calculation, we systematically decompose the mode equations into various regimes, depending on which term dominates in the dispersion relation, $\omega_k^2(\eta)$.  This approach has a broad applicability, beyond simply the spin-1 dark matter calculation that we have performed here. As a result, we find that the finite duration of reheating causes the spectrum of gravitationally-produced spin-1 particles to develop two breaks, associated with the scales that reenter the Hubble radius at the time when reheating ends and at the time when $m = H$; these results are summarized in \fref{fig:totalspectrum}.  Assuming that the spin-1 particles are stable and their comoving number density is conserved until today, we also calculate their relic abundance, which is shown in \fref{fig:omegavsm}.  For example, if $H_e \sim 10^{14} \GeV$ then the observed dark matter relic abundance is obtained if $m \sim 10^{-6} \eV$ and $50 \GeV \lesssim T_\mathrm{RH} \lesssim 10^{16} \GeV$ or if $T_\mathrm{RH} \sim 50 \GeV$ and $10^{-6} \eV \lesssim m \lesssim 10^{14} \GeV$.  To avoid producing too much dark matter, the parameters $H_e$, $T_\mathrm{RH}$, and $m$ are constrained, as shown in \fref{fig:constraint_plot}.  

In this work we have focused on understanding the gravitational production of vector dark matter during inflation and reheating. If this dark-matter candidate also has non-gravitational interactions, which simply did not play a role in its production, then a variety of observational probes become available, including direct detection in the lab.  On the other hand, if the dark matter only interacts with itself and visible matter through gravity, then observational prospects are clearly more challenging, but nevertheless several detection channels could be available.  Terrestrial probes, such as gravitational direct detection \cite{Carney:2019pza}, are most sensitive to larger dark photon masses; although, even for masses as large as $m \sim H_e \sim 10^{14} \GeV$, this signal would be very challenging to see.  Cosmological probes of spectator fields include isocurvature (between the dark matter and curvature perturbations) and non-Gaussianity (of the curvature perturbations).  Since the dark matter power spectrum is blue-tilted (falling toward smaller $k$) the isocurvature on CMB scales is predicted to be negligibly small~\cite{Graham:2015rva}.  On the other hand, in the quasi-single-field regime ($m \sim H_\mathrm{inf}$) the vector spectator may induce a detectable non-Gaussianity in the curvature perturbations \cite{Arkani-Hamed:2015bza,Wang:2020ioa} if it couples directly to the inflaton field.    Finally the blue-titled spectrum enhances the small-scale power in the dark matter perturbations, which may lead to the formation of primordial black holes \cite{Carr:2017edp} and provide additional astrophysical probes of this scenario.  

%==============
\section*{Acknowledgments}
The work of E.W.K. was supported in part by the US Department of Energy contract DE-FG02-13ER41958.  We are grateful to Evan McDonough for comments on the draft.  
%==============

%==============
%Bibliography
%==============
\bibliographystyle{JHEP}
\bibliography{vector}

%===================
\end{document}